\documentclass[
    aps,
    tightenlines, 
    prd, reprint, 
    groupedaddress,
    nofootinbib,
    amsmath,
    amssymb,
    amsfonts
]{revtex4-2}
\def\dd{{\rm d}}

\usepackage{graphicx}  
\usepackage{enumitem}
\usepackage{orcidlink}
\usepackage{dcolumn}  
\usepackage{bm}        
\usepackage{hyperref}  
\usepackage{soul}
\usepackage{empheq}

\hypersetup{
    colorlinks=true,
    linkcolor=red,   
    citecolor=blue,  
    urlcolor=blue,
}

\DeclareMathAlphabet{\mathdutchcal}{U}{dutchcal}{m}{n}
\SetMathAlphabet{\mathdutchcal}{bold}{U}{dutchcal}{b}{n}

\usepackage[dvipsnames]{xcolor}

\begin{document}

\title{Quantum black hole evaporation from thin shell collapse: a complete effective picture}

\author{Micha{\l} Bobula\orcidlink{0000-0002-4928-9568}}
\email{mbobula@fuw.edu.pl}
\affiliation{Faculty of Physics, University of Warsaw,
ulica Pasteura 5, 02-093 Warsaw, Poland}

\author{Francesco Fazzini\orcidlink{0009-0007-2263-5727}}
\email{francesco.fazzini@fau.de}
\affiliation{Institute for Quantum Gravity, Theoretical Physics III, Department of Physics,  Friedrich-Alexander-Universit\"at Erlangen-N\"urnberg, Staudtstraße 7, 91058 Erlangen, Germany}

\begin{abstract}
Quantum gravity is expected to play a central role in resolving classical singularities and providing a complete, {predictive} spacetime picture. In this paper, we develop an effective model inspired by loop quantum gravity that aims to fulfill this expectation, providing a unified description of thin shell collapse, black hole formation, evaporation, and remnant dynamics. The framework exhibits general validity, holding across different effective equations of motion, matter sources, and evaporation laws. {Specifically, we incorporate backreaction within a quasi-static approximation, modeling the radiation as null dust and matching the ingoing and outgoing fluxes across an evaporation surface using the Darmois-Israel junction conditions. For an evaporation law governed by the Stefan-Boltzmann law, we find that the final stage of evaporation yields an extremal black hole extending into a new asymptotic region, featuring both event and Cauchy horizons. Conversely, under the Schwinger approach, no event or Cauchy horizons form, instead, a temporary dynamical trapped region emerges, and the resulting spacetime remains globally hyperbolic. {Under this variant, our model predicts a horizonless null remnant escaping to future null infinity.}}

\end{abstract}
\maketitle

\section{Introduction}
 
Alongside spacetime singularities, event horizons may also be artifacts of General Relativity (GR). Defined as boundaries beyond which no signal can reach future null infinity \cite{Hawking:1973uf}, they are straightforward to identify in highly symmetric stationary black-hole spacetimes. In realistic dynamical settings, however, their location is determined by the global causal structure of the entire spacetime and is therefore intrinsically teleological. This has an important observational consequence: their existence cannot be established directly by experiment, since doing so would require complete knowledge of the spacetime's global causal structure \cite{Hayward:2000ca, Booth:2005qc, Visser:2014zqa, Ashtekar:2025qqa}.

This motivates the possibility that a quantum or ultraviolet-complete successor to GR may describe gravitational collapse and black-hole evolution without event horizons, while retaining only quasi-local structures such as trapping \cite{Hayward:1993wb} or apparent horizons \cite{Hawking:1973uf}. Unlike their global counterpart, these structures can, in principle, be identified through quasi-local observations and therefore provide a more operationally meaningful framework for black-hole phenomenology \cite{Visser:2014zqa, Ashtekar:2025qqa}.

Indeed, models of nonsingular (regular) black-hole formation followed by evaporation due to {Hawking radiation} can exhibit trapping horizons in the complete absence of event horizons \cite{Ashtekar:2005cj, Hayward:2005gi, Hossenfelder:2009fc, Frolov:2014jva, Frolov:2014wja, Frolov:2017rjz, Cardoso:2023guh, Gralla:2025gzl, DiFilippo:2025kzh, Boyanov:2025pes, Arrechea:2026hhl}, often forming a characteristic closed horizon trajectory in the spacetime \cite{Hayward:2005gi, Hossenfelder:2009fc, Frolov:2014jva, Frolov:2014wja, Frolov:2017rjz, Cardoso:2023guh, Gralla:2025gzl, DiFilippo:2025kzh, Boyanov:2025pes, Arrechea:2026hhl}. Such geometries are typically globally hyperbolic and thus avoid conceptual difficulties traditionally associated with black holes, including the information paradox \cite{Ashtekar:2005cj, Hayward:2005gi}. More importantly for phenomenology, the absence of a permanent causal boundary means that matter and radiation entering the trapped region may eventually return to the exterior. This opens, at least in principle, an observational window onto regions that would remain permanently inaccessible in the classical GR picture \cite{Frolov:2017rjz, Cardoso:2023guh, Arrechea:2026hhl}.

The natural question is whether this scenario can arise in a more robust quantum-corrected description of black-hole formation and evaporation. In recent years, considerable attention has been devoted to dynamical models of nonsingular black-hole formation \cite{Peleg:1995gg, Govender:2001gq, Bonanno:2006eu, Bojowald:2008ja, Bambi:2013caa, Zhang:2014bea, Schmitz:2019jct, Piechocki:2020bfo, Schmitz:2020vdr, Munch:2020czs, BenAchour:2020bdt, BenAchour:2020mgu, BenAchour:2020gon, Bonanno:2023rzk, Kelly:2020lec, Kelly:2020uwj, Husain:2021ojz, Husain:2022gwp, Giesel:2022rxi, Lewandowski:2022zce, Bobula:2023kbo, Fazzini:2023scu, Bobula:2024ywp, Bobula:2024chr, Cafaro:2024vrw, Shi:2024vki, Ou:2025bbv, Bobula:2026qzw, Giesel:2026pjj, Fazzini:2026uqp, Giesel:2025kdl, Han:2023wxg, Bobula:2024jlh, Cafaro:2026twe, Duque:2023syb, Bojowald:2024ium, Alonso-Bardaji:2023qgu, Malafarina:2022oka, Bueno:2025gjg, Li:2025kou}, although evaporation is still omitted in most of them. Representative examples are provided by approaches based on, or inspired by, leading candidates for a quantum theory of gravity, including loop quantum gravity (LQG) \cite{Husain:2021ojz, Lewandowski:2022zce, Giesel:2022rxi}, asymptotic safety \cite{Bonanno:2023rzk}, and string theory \cite{Bueno:2024eig}. A particularly important development is that these frameworks provide modified matter dynamics capable of describing the formation process and, in suitable settings, can be reformulated as effective cosmological dynamics for homogeneous and isotropic matter. This dynamical perspective goes beyond the earlier quantum-black-hole literature, which largely concentrated on stationary black-hole geometries, e.g. \cite{Ashtekar:2005qt, Gambini:2008dy, Gambini:2013ooa, Olmedo:2017lvt, Ashtekar:2018lag, Ashtekar:2018cay,  Simpson:2018tsi, Bodendorfer:2019cyv, Calza:2025mrt}.

Dynamical models of black-hole formation inspired by quantum-gravity approaches, typically formulated without backreaction, generally feature both event and Cauchy horizons \cite{Bonanno:2006eu, Schmitz:2020vdr, Munch:2020czs, BenAchour:2020gon, Bonanno:2023rzk, Kelly:2020lec, Kelly:2020uwj, Husain:2021ojz, Husain:2022gwp, Giesel:2022rxi, Lewandowski:2022zce, Bobula:2023kbo, Fazzini:2023scu, Bobula:2024jlh, Bobula:2024ywp, Bobula:2024chr, Cafaro:2024vrw, Shi:2024vki, Ou:2025bbv, Bobula:2026qzw, Giesel:2026pjj, Fazzini:2026uqp, Alonso-Bardaji:2023qgu, Bueno:2025gjg, Li:2025kou}. The latter present a serious obstruction, as they are subject to the mass-inflation instability \cite{Poisson:1988wc, Poisson:1989zz, Poisson:1990eh, Ori:1991zz}. Notably, in constructions that avoid Cauchy horizons, a consistent and successful embedding into cosmological dynamics is often lacking \cite{Simpson:2018tsi, Carballo-Rubio:2018jzw, Carballo-Rubio:2022kad, Ovalle:2023vvu, Zhang:2024ney, Calza:2025mrt}. This suggests that eliminating Cauchy horizons from quantum-corrected black-hole vacuum solutions may not be sufficient to resolve the underlying problems. Instead, as indicated by existing models \cite{Ashtekar:2005cj, Hayward:2005gi, Hossenfelder:2009fc, Frolov:2014jva, Frolov:2014wja, Frolov:2017rjz, Cardoso:2023guh, Gralla:2025gzl, DiFilippo:2025kzh, Boyanov:2025pes}, a more complete resolution may require the consistent inclusion of backreaction and the evaporation process. In this sense, backreaction may play a crucial role in preserving predictability and maintaining a consistent quantum-gravitational dynamics.

In this work, we focus on quantum-corrected dynamics derived from symmetry-reduced frameworks inspired by LQG \cite{ Kelly:2020lec, Kelly:2020uwj, Husain:2021ojz, Husain:2022gwp, Giesel:2022rxi, Lewandowski:2022zce, Bobula:2023kbo, Fazzini:2023scu, Bobula:2024ywp, Bobula:2024chr, Cafaro:2024vrw, Shi:2024vki, Ou:2025bbv, Bobula:2026qzw, Giesel:2025kdl, Giesel:2026pjj, Fazzini:2026uqp} and extend these models to incorporate backreaction. For spherically symmetric dust collapse, the quantum corrections resolve the central singularity through a quantum bounce, with the matter emerging into a new asymptotic region and the resulting causal structure resembling that of a Reissner--Nordström black hole. A common feature of these works \cite{ Kelly:2020lec, Kelly:2020uwj, Husain:2021ojz, Husain:2022gwp, Giesel:2022rxi, Lewandowski:2022zce, Bobula:2023kbo, Fazzini:2023scu, Bobula:2024ywp, Bobula:2024chr, Cafaro:2024vrw, Shi:2024vki, Ou:2025bbv, Bobula:2026qzw, Giesel:2025kdl, Giesel:2026pjj, Fazzini:2026uqp} is that they identify the same effective, spherically symmetric quantum-corrected black-hole vacuum geometry, characterized by the Schwarzschild-like metric function $f_\mathrm{qS} (r) :=1- \frac{2 GM}{r} + \frac{\gamma^2 \Delta(2 GM)^2 }{r^4}$, where $\gamma$ is the Barbero-Immirzi parameter and $\Delta := 4 \sqrt{3} \pi \gamma \ell_\mathrm{Pl}^2$ is the so-called \emph{area gap}, i.e., the smallest nonzero eigenvalue of the area operator in LQG, with $\ell_\mathrm{Pl}=\sqrt{G\hbar}$ denoting the Planck length. In particular, Refs. \cite{Lewandowski:2022zce, Bobula:2023kbo} showed that this geometry is uniquely selected by requiring consistency with the effective dust dynamics of the Ashtekar-Paw{\l}owski-Singh cosmological model \cite{Ashtekar:2006rx, Ashtekar:2006uz, Ashtekar:2006wn, Husain:2011tm}.

To model backreaction, we follow the approach of Hiscock \cite{Hiscock:1980ze, Hiscock:1981xb}, in which {Hawking radiation} is described by ingoing Vaidya-like geometries (representing Hawking-like partners) and outgoing Vaidya-like geometries (representing Hawking-like particles), matched across a surface near the trapping (evaporating) horizon through the first Darmois-Israel junction condition. We estimate the corresponding flux and subsequently obtain time-dependent geometry via quasi-static approach using two different prescriptions: (i) Stefan-Boltzmann black-body emission \cite{Page:1976df} and (ii) a Schwinger-type approach to {Hawking-like evaporation.} \cite{Wondrak:2023zdi, Falcke:2024rdu}\footnote{It remains an open question whether the Schwinger-type particle production presented in \cite{Wondrak:2023zdi, Falcke:2024rdu} is physically identical to the mechanism in Hawking's seminal work \cite{Hawking:1975vcx}. Nevertheless, their 
similarity  suggests they might be manifestations of the same underlying phenomenon, therefore, we adopt the term 'Hawking-like' to describe it.}. To obtain a complete spacetime picture of matter collapse, black-hole formation, and evaporation, we consider the collapse of a timelike thin dust shell to form the black hole, using the Hamiltonian formalism for junction conditions, describing the thin shell's interaction with ingoing negative-energy null dust. We gain analytic control over the null-dust dynamics by extending the timelike dust framework used in LQG-inspired models to the null case via Lemos' trick \cite{Lemos1992}, namely by taking an appropriate limit of the binding function appearing in the dust dynamics.

Within the considered dynamical framework, we find that a consistent, globally hyperbolic description of black-hole formation and evaporation is achievable. The resulting spacetime features the characteristic closed trajectory of the trapping horizons and demonstrates robustness against variations in model parameters. Crucially, the model predicts the emission of a Planck-mass ($\sim m_{\mathrm{Pl}}$) remnant null shell  expanding toward an asymptotically flat region---which is the same region of the collapsed thin shell in the Schwinger approach, while a new asymptotic region in the Stefan-Boltzmann approach---once the trapped region disappears.
We identify this shell as a Planck-mass messenger carrying quantum gravitational effects arising from our LQG-inspired framework.

{This work is organized as follows. In Sec.~\ref{sec_lqgdustcollapse}, we review the effective dynamics of timelike dust within an LQG-inspired framework, which we use to describe the black hole formation stage. In Sec.~\ref{sec_nulldust}, we extend this framework to incorporate null dust via Lemos' trick in order to model Hawking-like radiation. In Sec. \ref{sec_evaporationlaws}, we present two evaporation laws that quantify the outgoing null dust radiation. In Sec. \ref{sec_hawking_israel}, we match the outgoing and ingoing null-dust spacetimes across an evaporation surface using the Israel–Darmois junction conditions. In Sec. \ref{secdynamicsthin}, we formulate the dynamics of the thin shell forming the black hole as it interacts with the ingoing null dust representing Hawking-like partners. In Sec. \ref{sec_completespacetime}, we construct the complete spacetime pictures for the considered framework and evaporation laws, presenting their corresponding conformal diagrams. {In Sec. \ref{sec_comparison} we compare our results with earlier literature.} Finally, in Sec. \ref{sec_discussion}, we discuss our {findings} and conclude.  }

\section{LQG-inspired dust collapse} \label{sec_lqgdustcollapse}
To construct our model of an evaporating black hole formed through the collapse of a timelike thin shell, with the Hawking-like flux represented by null dust, we first recall the canonical framework for spherically symmetric gravity coupled to dust in LQG-inspired models \cite{ Kelly:2020lec, Kelly:2020uwj, Husain:2021ojz, Husain:2022gwp, Giesel:2022rxi, Lewandowski:2022zce, Bobula:2023kbo, Fazzini:2023scu, Bobula:2024ywp, Bobula:2024chr, Cafaro:2024vrw, Shi:2024vki, Ou:2025bbv, Bobula:2026qzw, Giesel:2025kdl, Giesel:2026pjj, Fazzini:2026uqp}. In Sec. \ref{sec_sub_coupledtodust}, we formulate the canonical system of a dust field coupled to classical gravity in spherical symmetry using the dust time gauge. Sec. \ref{sec_sub_LTB} then focuses on the Lemaître--Tolman--Bondi (LTB) gauge prior to discussing the effective LQG-inspired dynamics. In Sec. \ref{sub_sec_arealgauge}, we introduce the areal gauge as an alternative to the LTB gauge, which is employed complementarily throughout this work. Additionally, we examine the connection between the presented effective dynamics and existing quantum descriptions \cite{Husain:2021ojz, Husain:2022gwp, Cafaro:2026twe}.

\subsection{Dust-coupled gravity in spherical symmetry and the dust time gauge } \label{sec_sub_coupledtodust}
The starting point is the action for general relativity coupled to pressureless dust,
\begin{equation}
S=\frac{1}{16 \pi G} \int \mathrm{d}y \sqrt{-\det(g)}
\left[
\mathcal{R}
-\frac{\rho}{2}
\left(
g^{\mu\nu}T_{;\mu}T_{;\nu}+1
\right)
\right],
\end{equation}
where $\mathcal{R}$ denotes the four-dimensional Ricci scalar, $g_{\mu\nu}$ is the spacetime metric, $\rho$ is the dust energy density, and $T$ is the field defining the dust flow. Performing an Arnowitt-Deser-Misner (ADM) decomposition and subsequently rewriting the gravitational phase space in terms of densitized triads and Ashtekar--Barbero connections provides a convenient starting point for the LQG treatment. Upon imposing spherical symmetry, the gravitational phase space is parametrized by two independent components of the densitized triad, $E^x$ and $E^\varphi$, together with their canonically conjugate connection variables represented by extrinsic curvature components, $K_x$ and $K_\varphi$, respectively \cite{Giesel:2023tsj}. Their nonvanishing Poisson brackets are
\begin{equation}
\begin{aligned}
\{K_x(x,t),E^x(x',t)\}
&=G \delta(x-x')\,, \\
\{K_\varphi(x,t),E^\varphi(x',t)\}
&=G \delta(x-x')\,.
\end{aligned}
\end{equation}
In terms of the triad variables, the spherically symmetric spacetime metric takes the form
\begin{equation} \label{adm_reduced}
\mathrm{d}s^2
=-N^2\mathrm{d}t^2
+\frac{(E^\varphi)^2}{|E^x|}
\left(\mathrm{d}x+N^x\mathrm{d}t\right)^2
+|E^x|\mathrm{d}\Omega^2 \,,
\end{equation}
where $N$ and $N^x$ denote the lapse and radial shift, respectively, and $|E^x|\dd\Omega^2$ is the metric on the two-sphere. The triad component $E^x$ determines the areal radius $r$ via $r=\sqrt{|E^x|}$, while $E^\varphi$ encodes the radial component of the spatial geometry. The ADM action in these variables takes the form
\begin{equation}
\begin{aligned} 
S = & \int \mathrm{d}t \, \mathrm{d}x \, \Big[  \frac{E^x \partial_t K_x +  E^{\varphi}  \partial_t K_\varphi} { 2G}  + 4 \pi p_T \partial_t T \\ 
& - N \left(\mathcal{H}^\mathrm{grav} + \mathcal{H}^\mathrm{matt}\right) - N^x \left(\mathcal{D}^\mathrm{grav} + \mathcal{D}^\mathrm{matt}\right) \Big] \,,
\end{aligned}
\end{equation}
where $ \mathcal{H}^\mathrm{grav} + \mathcal{H}^\mathrm{matt} $ and $ \mathcal{D}^\mathrm{grav} + \mathcal{D}^\mathrm{matt}  $ denote the total Hamiltonian and diffeomorphism constraints, respectively, incorporating both gravitational and matter contributions. The $p_T$ is the dust momentum canonically conjugate to $T$.

We next choose the dust proper time as the time coordinate by fixing the foliation of spacetime according to $T-t=0$. This choice identifies the coordinate time $t$ with the proper time measured by comoving dust observers and, in the corresponding gauge, fixes the lapse to $N=1$.

\subsection{LTB gauge} \label{sec_sub_LTB}

To recover the LTB form of the metric, the two triad components $E^x$ and $E^\varphi$ cannot be treated as independent. Instead, they are related by the LTB condition $
\left|E^x\right|^{\prime}
-2E^\varphi\sqrt{1+\varepsilon(x)}=0$ \cite{Giesel:2023tsj, Giesel:2024mps, Cafaro:2026twe}, 
where a prime denotes differentiation with respect to the radial coordinate $x$, and $\varepsilon(x)$ is the LTB energy function. This condition ensures that $x$ serves as a comoving LTB coordinate, remaining constant along the worldlines of the timelike dust particles. The dust-time gauge fixes the time-reparametrization freedom generated by the Hamiltonian constraint, whereas the LTB condition fixes the radial diffeomorphism freedom generated by the diffeomorphism constraint. With this steps, the metric takes the form 
\begin{equation}
\mathrm{d} s^2=-\mathrm{d} t^2+\frac{\left(\partial_x E^x\right)^2}{4\left|E^x\right|(1+\varepsilon(x))} \mathrm{d} x^2+\left|E^x\right| \mathrm{d} \Omega^2 \,,
\label{ltbform}
\end{equation}
{where clearly we have $N^x=0$}.
The gravitational part of the Hamiltonian constraint then takes the form \cite{ Giesel:2025kdl, Fazzini:2026uqp}
\begin{equation}
\mathcal{H}^\mathrm{grav}
=-\frac{\partial_x \tilde{H}_{\mathrm{GR}}}
{\sqrt{1+\varepsilon(x)}}\,,
\end{equation}
where $\tilde{H}_{\mathrm{GR}}(x)$ denotes a shell Hamiltonian, that is a physical Hamiltonian generating dynamics on a phase space reduced to single dust shell of constant $x$. In the present formulation, it coincides with the Misner–Sharp mass $M(x)$ enclosed within the shell, such that $\tilde{H}_{\mathrm{GR}}(x)=M(x)$. One can further demonstrate that this classical setup leads directly to the classical LTB equation of motion $(\partial_t r)^2 = \varepsilon(x) + 2 GM(x)/r $.

Throughout this work, we focus on effective LTB-like dynamical equations of the form
\begin{equation} \label{ltb_general}
 \left( \partial_t r \right)^2=    \varepsilon+r^2 \tilde{f}\left(\frac{2 G M}{r^3}\right)\,,
\end{equation}
where the function $\tilde{f}(X)$ encodes deviations from the classical GR dynamics, recovered for $\tilde{f}(X)=X$. {Many well-known regular black hole spacetimes, such as the Hayward and Bardeen geometries, admit a dynamical formation described by \eqref{ltb_general} \cite{Giesel:2025kdl, Ling:2026kjp}.} In particular, we are interested in effective dynamics for which the $f_\mathrm{qS}(r)$ black-hole geometry arises as the unique vacuum solution. The quantum-corrected vacuum black-hole geometry \cite{ Kelly:2020lec, Kelly:2020uwj, Husain:2021ojz, Husain:2022gwp, Giesel:2022rxi, Lewandowski:2022zce, Bobula:2023kbo, Fazzini:2023scu, Bobula:2024ywp, Bobula:2024chr, Cafaro:2024vrw, Shi:2024vki, Ou:2025bbv, Bobula:2026qzw, Giesel:2025kdl, Giesel:2026pjj, Fazzini:2026uqp} is obtained as the vacuum solution for the choice  $
\tilde{f} \left( X \right) =  X \left[ 1 - \gamma^2 \sqrt{\Delta} X  \right]$
with the corresponding metric function given by \cite{Giesel:2025kdl}
\begin{equation}
 1-  r^2 \tilde{f}\left( \frac{2 G M}{r^3} \right) \to  1- \frac{2 GM}{r} + \frac{\gamma^2 \Delta(2 GM)^2 }{r^4} = f_\mathrm{qS}(r)\,.
 \end{equation}
Then the effective LTB-like Eq. \eqref{ltb_general} becomes
\begin{equation} \label{ltb}
 \left( \partial_t r \right)^2=    \varepsilon+ \frac{2GM}{r}\left(1-\frac{2GM\gamma^2 \Delta}{r^3} \right) \,.
\end{equation}

To relate the formalism to the LQG-inspired effective dynamical schemes presented in \cite{Husain:2021ojz, Husain:2022gwp, Giesel:2025kdl, Cafaro:2026twe}, we introduce variables $\mathdutchcal{v}:= (E^x)^{3/2}$, $b:= K_\varphi /\sqrt{E^x}$. Then the shell Hamiltonian in classical GR takes the form 
\begin{equation}
\label{Hclassical}
\tilde{H}_\mathrm{GR} = \frac{1}{2G}\left(b^2 \mathdutchcal{v}
-\varepsilon v^{1/3}\right) \,.
\end{equation}
On the reduced single-shell phase space, the variables satisfy the non-distributional Poisson bracket $\{ v, b \} =  - {3} G$. The effective LTB-like Eq. \eqref{ltb} corresponds to the unique shell Hamiltonian encoding LQG-inspired corrections to the dynamics \cite{Giesel:2024mps} 
\begin{equation} \label{hamiltonian_shell}
\begin{aligned}
\tilde{H}_\mathrm{eff}=\frac{\mathdutchcal{v}}{2G} & {\left[\frac{1}{\gamma^2 \Delta } \sin ^2\left(\gamma \sqrt{\Delta} b+\frac{\gamma \sqrt{\Delta \varepsilon}}{\mathdutchcal{v}^{1 / 3}}\right)\right.} \\
& \left.-\frac{\sqrt{\varepsilon}}{\gamma \sqrt{\Delta} \mathdutchcal{v}^{1 / 3}} \sin \left(2 \gamma  \sqrt{\Delta} b+\frac{2 \gamma \sqrt{\Delta \varepsilon}}{\mathdutchcal{v}^{1 / 3}}\right)\right] ~.
\end{aligned}
\end{equation}
In the marginally bound limit, $\varepsilon = 0$, the Hamiltonian simplifies to
\begin{equation}
  \tilde{H}_\mathrm{eff} = \frac{\mathdutchcal{v}}{2 G} \frac{1}{(\gamma \Delta)^2  } \sin^2 \left( \gamma \Delta b \right) \,,
\end{equation}
recovering the effective shell Hamiltonian derived in \cite{Cafaro:2026twe}.

We clarify the relation between dynamical Eq. \eqref{ltb} and the quantization procedure developed in \cite{Cafaro:2026twe, Husain:2021ojz, Husain:2022gwp}. These works loop quantize the dust system and derive effective dynamics via an Ehrenfest-like semiclassical approximation. However, the resulting effective Hamiltonian reproduces Eq. \eqref{hamiltonian_shell} only for marginally bound collapse, reflecting the absence of a Birkhoff-like property: as noted in \cite{Giesel:2025kdl}, the vacuum solution depends on $\varepsilon$. This issue was addressed at the effective level in \cite{Giesel:2025kdl}, which proposes the effective Hamiltonian \eqref{hamiltonian_shell} that is unique under reasonable assumptions, though without developing a corresponding quantum description.

We emphasize that the derivations presented in the following Secs. \ref{sec_nulldust}, \ref{secdynamicsthin} hold for arbitrary LTB-like dynamical equations in the class of \eqref{ltb}, rather than being restricted to the specific choice of $\tilde{f}$ present in Eq. \eqref{ltb}. Although specific metric functions—such as the $f_\mathrm{qS}(r)$—are employed to reach the analytical and numerical results, the generality of the framework is preserved.

\subsection{Areal gauge} \label{sub_sec_arealgauge}
For the subsequent computations in this work, we also introduce an alternative spatial gauge choice, namely the areal gauge, defined by $|E^x|=x^2$, so that the coordinate $x$ coincides with the areal radius. To avoid notational ambiguity, however, we continue to denote the areal radius by $r$. In this gauge, the metric \eqref{adm_reduced} takes the form
\begin{equation} \label{adm_reduced}
\mathrm{d}s^2
=-\mathrm{d}t^2
+\frac{(E^\varphi)^2}{r^2}
\left(\mathrm{d}r+N^r\mathrm{d}t\right)^2
+r^2\mathrm{d}\Omega^2 \,.
\end{equation}
This chart describes the dust dynamics in Painlev\'e-Gullstrand-like (PG-like) coordinates, provided the triad angular component takes the form of $E^\varphi= r/\sqrt{1 + \varepsilon(r,t)}$. It can be checked that this is the general solution for the {equations of motion} of $E^\varphi$ in PG-like coordinates \cite{Cipriani:2024nhx}.
The radial shift \(N^r\) is kinematically related to the areal radius through \(N^r=-\partial_t r\), where the exact expression follows from Eq. \eqref{ltb_general}.

\section{Null dust and effective Vaidya spacetime} \label{sec_nulldust}
The effective equations introduced in the previous section describe the evolution of dust in marginal, non-marginal and bound configurations. Since, together with the thin shell dynamics, we want to study the backreaction of {Hawking-like radiation} (both particles and partners) to the geometry using the Hiscock approach, where such energy fluxes are modeled by null dust, we need to extend the model to the null case. 
Studying null dust in general requires null coordinates, and the Hamiltonian formulation in null coordinates is notoriously challenging already at the classical level. Instead of facing this issue directly, we can buy a strategy that turns out to be successful in the classical theory, and is mainly due to Lemos \cite{Lemos1992}, and extended by Hellaby     \cite{Hellaby:1994bj}. The basic idea is to introduce null dust as a limit of the massive dust in the unbound configuration, and check a posteriori the crucial features required by null dust.

Here we will implement directly Lemos' strategy for the effective model, while refer to literature \cite{Hellaby:1994bj} for the classical counterpart. In order to follow Lemos' formulation, we need to cast the effective LTB metric in PG-like coordinates. To do this, we perform the following change of coordinates to \eqref{ltbform}: $r=r(x,t)$, which leads to
\begin{equation} \label{metric1} 
\begin{aligned}
\dd s^2 =-\dd t^2 & +\frac{1}{1+\varepsilon(r,t)}\left(\dd r + N^r \dd t \right)^2+r^2 \dd \Omega^2 \,, \\
 N^r = & \pm {\sqrt{\varepsilon(r,t)+ r^2\tilde{f}\left( \frac{2 G M(r,t) )}{r^3} \right)} }\,,
\end{aligned}
\end{equation}
where $\pm$ determines a collapsing/expanding dust solution, and where
{ the radial shift corresponding to the metric function $f_\mathrm{qS}(r)$ and the specific LTB-like dynamics \eqref{ltb} takes the form}
\begin{equation}  \label{shift}
N^r =\pm \sqrt{\varepsilon(r,t)+\frac{2GM(r,t)}{r}\left(1-\frac{2GM(r,t)\gamma^2 \Delta}{r^3} \right)} \,.
\end{equation}
$M(r,t)$ and $\varepsilon(r,t)$ follow an effective version of the classical equations of motion. The reason why such coordinates are called {PG-like} is that while the time $t$ is the dust proper time, the spatial metric is not flat. 

We can set $\varepsilon(r,t)=\mathrm{const}.$ This is indeed a solution of the {equations of motion} for $\varepsilon$ in these coordinates. To check this, we can start from the conservation of $\varepsilon$ in LTB form, and perform the change of coordinates from LTB
\begin{align}
 \partial_t^\mathrm{LTB} \varepsilon({x}) = \partial_t^\mathrm{PG} &\varepsilon(r,t)+\partial_r \varepsilon(r,t)\partial_t r= \notag \\ 
 &=\partial_t^\mathrm{PG} \varepsilon(r,t)-  {N^r } \partial_r \varepsilon=0 \,,
\end{align}
where the suffix specifies that even though $t$ is the same for the two charts (the dust proper time), the components $\partial_t$ are not. One can immediately check that $\varepsilon(r,t)=\mathrm{const.}>0$ is solution of the previous {partial differential equation}. Clearly, it is not the most general solution, but it is required for our aims.

We can then write more explicitly the line element
\begin{align} \label{met_eps}
 \dd s^2=& -\frac{1}{1+\varepsilon}\left[ {1-r^2\tilde{f}\left( \frac{2 G M(r,t) )}{r^3} \right) } \right]\dd t^2 +\frac{\dd r^2}{1+\varepsilon}+ \notag \\
 &+\frac{2}{1+\varepsilon}{N^r}\dd r \dd t + r^2 \dd \Omega^2  ~.
\end{align}
This metric, as said before, describes non-marginally bound time-like dust collapse in the effective model.

Lemos' trick can be applied both to expanding and collapsing massive dust.
In the collapse case ($N^r>0$), it consists in 
performing the following linear change of coordinates 
\begin{equation}
v=\frac{t}{\sqrt{\varepsilon}}+\frac{r}{\varepsilon}    \,, \label{v} 
\end{equation}
and then send $\varepsilon\rightarrow +\infty$,
which leads, after a brief computation, to
\begin{equation} \label{vaidya}
\dd s^2 =-\left[{1-r^2\tilde{f}\left( \frac{2 G M(v, r) )}{r^3} \right) } \right]\dd v^2 +2 \dd r \dd v +r^2 \dd \Omega^2 \,,
\end{equation}
 which is (almost) the effective version of the Vaidya metric in advanced Eddington-Finkelstein coordinates. Note that sending $\varepsilon\rightarrow +\infty$ is not a change of coordinates, as would be in the case of the Schwarzschild metric \cite{Martel:2000rn}, but it is a limit performed in the space of physical solutions. To get an effective Vaidya metric, we need to prove that $M(v,r)=M(v)$. The reason for this comes from the physical interpretation of the $v$ coordinate. As in the classical Schwarzschild and Vaidya cases, it can be easily checked that $v=\mathrm{const}.$ identifies ingoing radial null geodesics of the effective spacetime. Since here we look at ingoing null dust (which since the absence of internal pressure must follow null geodesics), then the mass enclosed by a certain null shell $v$ should be a function of $v$ only, due to the fact that different shells are not interacting (no energy exchange due to the absence of radial pressure). The logic here is exactly the same as in the LTB massive dust case, where the dust shells are parametrized by $x$, and the mass depends only on the $x$ coordinate.

In order to prove that $M=M(v)$, we can look at the LTB formulation of the metric. In the limit $\varepsilon \rightarrow +\infty$, Eq. \eqref{ltb_general} reads

\begin{equation} 
 \partial_t r(x,t)\rightarrow -\sqrt{\varepsilon} ~,  \label{vellim}
\end{equation}
which has, as collapsing solution: $r(x,t)\rightarrow \sqrt{\varepsilon}(a(x)-t)$, as in the classical case.
In this limit, the $v$ coordinate is related with the comoving radius through

\begin{equation}
 v=\frac{a(x)}{\sqrt{\varepsilon}}\,,   
\end{equation}
as one can immediately check by plugging the solution $r(x,t)$ in \eqref{v}. This can be seen as a coordinate transformation between the LTB and the advanced Eddington-Finkelstein coordinates, and since the mass function in LTB coordinates depends only on $x$, then will automatically depend only on $v$ in any chart that contains the $v$ coordinate, as the Eddington-Finkelstein one $\{v,r,\theta, \phi \}$. This leads to a complete identification of \eqref{vaidya} with the effective version of the Vaidya metric in advanced Eddington-Finkelstein coordinates. 

Let's give a physical interpretation of this result. In PG-like coordinates, the $4$-velocity of the dust field is given by

\begin{equation} \label{uvec}
 u^\mu=\left(1, \frac{\dd r}{\dd t},0,0\right)=(1,-N^r,0,0)~, 
\end{equation}
where the last equality can be easily checked by solving $u^\mu u_\mu =-1$.
Now, in the asymptotic limit $r\rightarrow +\infty$, assuming a bounded mass function (realized by an initial density profile of non-compact support), the PG-like metric reads

\begin{equation}
 \dd s^2=-\frac{\dd t^2}{1+\varepsilon}+\frac{\dd r^2}{1+\varepsilon}+\frac{2\sqrt{\varepsilon}}{1+\varepsilon}\dd r \dd t +r^2 \dd \Omega^2~.   
 \label{minkopg}
\end{equation}
We can cast this metric in Minkowski form, with the change of coordinates

\begin{equation}
  t=\sqrt{\varepsilon}~r+ \sqrt{1+\varepsilon}~t_{M}~.  
\end{equation}
One can then write the dust three-velocity at infinity \eqref{vellim} in Minkowski time, leading to, after a brief computation

\begin{equation}
 \frac{\dd r}{\dd t_M}=-\frac{\sqrt{\varepsilon}}{\sqrt{\varepsilon+1}} ~. 
\end{equation}
This means that $\varepsilon\rightarrow +\infty$ leads from the unbound massive dust solution ($0<\varepsilon<+\infty$ finite) to the massless case, for which at infinity ${\dd r}/{\dd t_M}=-1$, and explains why one recovers the Vaidya solution, which describes null dust collapse, in such a limit. The reason why \eqref{vellim} cannot be used directly for the comparison with the null case is that the PG-like asymptotic metric does not coincide with the Minkowski form, so in the PG-like chart the radial asymptotic speed of light is not $\pm 1$ (in the adopted units $c=1$). 

The derivation of the outgoing effective Vaidya solution follows exactly the same steps described above, with the only difference of starting with outgoing massive dust, that means a negative shift vector in {PG-like} coordinates. The resulting spacetime
\begin{equation}  \label{vaidyaout}
    \dd s^2 =-\left[{1-r^2\tilde{f}\left( \frac{2 G M(u) )}{r^3} \right) } \right]\dd u^2 -2 \dd r \dd u +r^2 \dd \Omega^2
    \,,
\end{equation}
where, as in the classical case, $u$ labels outgoing radial null dust shells.

A last comment is in order. In principle one can recover the same result with a general $\varepsilon(r,t)$. However, the coordinate transformation from $t$ to $v$ becomes more involved, and since the non-marginally bound state of dust is only an intermediate step in our procedure, we will not deal with this complication. Also, as noted in \cite{Hellaby:1994bj}, a constant $\varepsilon$ would not allow to recover the Vaidya spacetime if the dust fills the complete spacetime. This can be easily understood by the fact that the origin of the coordinates should be described by a stationary shell $\partial_t r(R,t)=0$, which, in LTB collapse, is allowed by the condition $M(R=0)=0$, $\varepsilon(R=0)=0$. This is clearly not possible if $\varepsilon$ is a non-zero constant. However, we will see that the assumption of constant $\varepsilon$ in our model is completely consistent, since the dynamics does not allow the dust-filled region (null or timelike) to reach $r=0$, since such a region is radially bounded from below by an evolving bouncing thin shell that cannot reach $r=0$. We will come back to this issue later, while for a more extensive discussion, see \cite{Hellaby:1994bj}.
\section{Evaporation laws} \label{sec_evaporationlaws}
Effective Vaidya spacetimes provide a useful framework for modeling the geometric backreaction of Hawking pair production. However, to describe Hawking-like evaporation, one must prescribe the energy flux carried by the outgoing positive-energy radiation (Hawking-like particles). In the quasi-static approximation employed here, the black-hole geometry is treated as evolving through a sequence of instantaneously stationary states: while the mass parameter $M(u)$ decreases slowly over time, the outgoing flux at any given moment is estimated using a momentarily static background corresponding to that instantaneous mass. In this section, we consider two outgoing flux estimates from the literature, one older and one more recent, which lead to different quantitative and qualitative predictions. {The results are summarized in Fig. \ref{fig_MM_comparison}, where the $M(u)$ functions are plotted.} In the next Sec. \ref{sec_hawking_israel}, we employ the Darmois–Israel junction conditions to derive the evaporation law governing the backreaction of the ingoing negative-energy flux carried by the Hawking-like partners.

\subsection{Stefan-Boltzmann approach}
The evaporation law we consider here can be derived from the Stefan-Boltzmann law applied to the black hole, in the approximation of considering it as a perfect black body with temperature given by the effective Hawking temperature \cite{Bonanno:2006eu, Hossenfelder:2009fc, Bonanno:2022jjp}. 

 The starting point is to consider the Hawking temperature defined through the surface gravity of the outer {trapping} horizon of a static black hole of consideration. A simple computation leads to the following
\begin{equation}
\mathcal{T}_\mathrm{Hawk}=\frac{\hbar \kappa}{2 \pi k_B} = \frac{\hbar \partial_rf(r)|_{r=r_\mathrm{out}}}{4 \pi k_B} \,,
\end{equation}
where $k_B$ is Boltzmann constant and $\kappa$ is the surface gravity evaluated at the outer horizon, $r=r_\mathrm{out}$, operationally given as the largest positive root of $f(r):=1-r^2\tilde{f}\left(\frac{2GM}{r^3}\right)$ where $f$ is the metric function of a static Schwarzchild-like line element. Then, the Stefan–Boltzmann law gives the luminosity (energy flux integrated over the emitting surface) 
\begin{equation}
    \frac{\partial E}{\partial u}=4\pi r_\mathrm{out}^2 \sigma_{\mathrm{SB}} \mathcal{T}_H^4 \,,
\end{equation}
where $\sigma_{\mathrm{SB}}:= \frac{\pi^2 k_B^4 }{60 \hbar^3}$.
Within the quasi-static approximation to backreaction, energy conservation requires the emitted energy to be balanced by the loss of black hole mass, so that
\begin{equation} \label{Boltz}
    \frac{\partial M}{\partial u}=-4\pi r_H^2 \sigma_{\mathrm{SB}} \mathcal{T}_H^4 \,.
\end{equation}
Hence, Eq. \eqref{Boltz} becomes autonomous ordinary differential equation whose solution is time-dependent mass $M(u)$. For macroscopic black holes $\sqrt{\Delta}/2GM\ll 1$ governed by the metric function $f_\mathrm{qS}(r)$ we get
\begin{equation}
    \frac{\partial M}{\partial u}=O\left( \frac{\sqrt{\Delta}}{2 G M}  \right)^2\,.
\end{equation}
For black holes featuring both an inner and an outer horizon, the evaporation law in Eq. \eqref{Boltz} typically yields a mass function $M(u)$ that asymptotically approaches an extremal mass value $M_\mathrm{ext}$. At this limit, the two horizons merge into a single degenerate horizon as the temperature vanishes. Consequently, this picture generally leaves behind a Cauchy horizon once the evaporation process asymptotically terminates.

\subsection{Schwinger approach}
 We also consider an alternative approach to estimating the emitted energy flux based on curvature-induced particle production \cite{Wondrak:2023zdi, Falcke:2024rdu}. This approach treats the real massless scalar quantum field as propagating on a prescribed curved-spacetime background, without requiring the gravitational field equations to be imposed. Using heat-kernel methods and covariant perturbation theory, the authors derive an expression for the imaginary part of the one-loop effective action, which they interpret as a measure of gravitationally induced particle production and use it to estimate the associated energy flux for the Schwarzschild black hole. In present work, we will use this approach for calculate the flux for the quantum-corrected black hole of our consideration. The imaginary part of the effective action is
\begin{widetext}
\begin{equation} \label{effective_action}
\operatorname{Im}(W)=  \int \mathrm{d} y \sqrt{-g} \, \operatorname{Im} \left( \mathcal{L}_\mathrm{} \right) =  \frac{ \hbar }{64 \pi} \int \dd y \sqrt{-g}\left[\frac{1}{2}\left(\xi-\frac{1}{6}\right) \mathcal{R}^2\right.
 \left.+\frac{1}{180}\left(\mathcal{R}_{\mu \nu \rho \sigma} \mathcal{R}^{\mu \nu \rho \sigma}-\mathcal{R}_{\mu \nu} \mathcal{R}^{\mu \nu}\right)\right] \,,
\end{equation}
\end{widetext}
which includes contractions of the Riemann tensor $\mathcal{R}^{\mu\nu\sigma\rho}$ and the Ricci tensor $\mathcal{R}^{\mu\nu}$. In Eq. \eqref{effective_action}, $\xi$ is the gravitational coupling parameter, where we set $\xi = 1/6$ to represent conformal coupling—viable, for instance, for a single polarization state of a photon \cite{Wondrak:2023zdi, Falcke:2024rdu}.

For metrics of the form given in Eq. \eqref{vaidyaout}, we can use Eq. \eqref{effective_action} to express the imaginary part of the effective Lagrangian as
\begin{equation}
\operatorname{Im}(L_\mathrm{}) = 4 \pi \int_{r_*}^\infty \mathrm{d} r \, r^2 \operatorname{Im}(\mathcal{L}_\mathrm{})\,~,
\end{equation}
where the integration extends from an arbitrary radius $r=r_*$ to $\infty$, as we are interested in particle production within this as-yet-unspecified domain. This yields the particle production rate density with respect to $u$ time $\frac{ \mathrm{d} N_{\mathrm{e}}}{\mathrm{d} u} = \frac{2}{\hbar} \operatorname{Im} (L_\mathrm{eff})$. We can also write the radial profile of the pair-production event rate as follows
\begin{equation} \label{dNdudr}
\frac{\mathrm{d}^2 N_{\mathrm{e}}}{ \mathrm{d}u  \mathrm{d}r} = \frac{ 8 \pi r^2}{\hbar} \operatorname{Im} (\mathcal{L}_\mathrm{})\,.
\end{equation}
\begin{figure}
    \centering
    \includegraphics[width=1\linewidth]{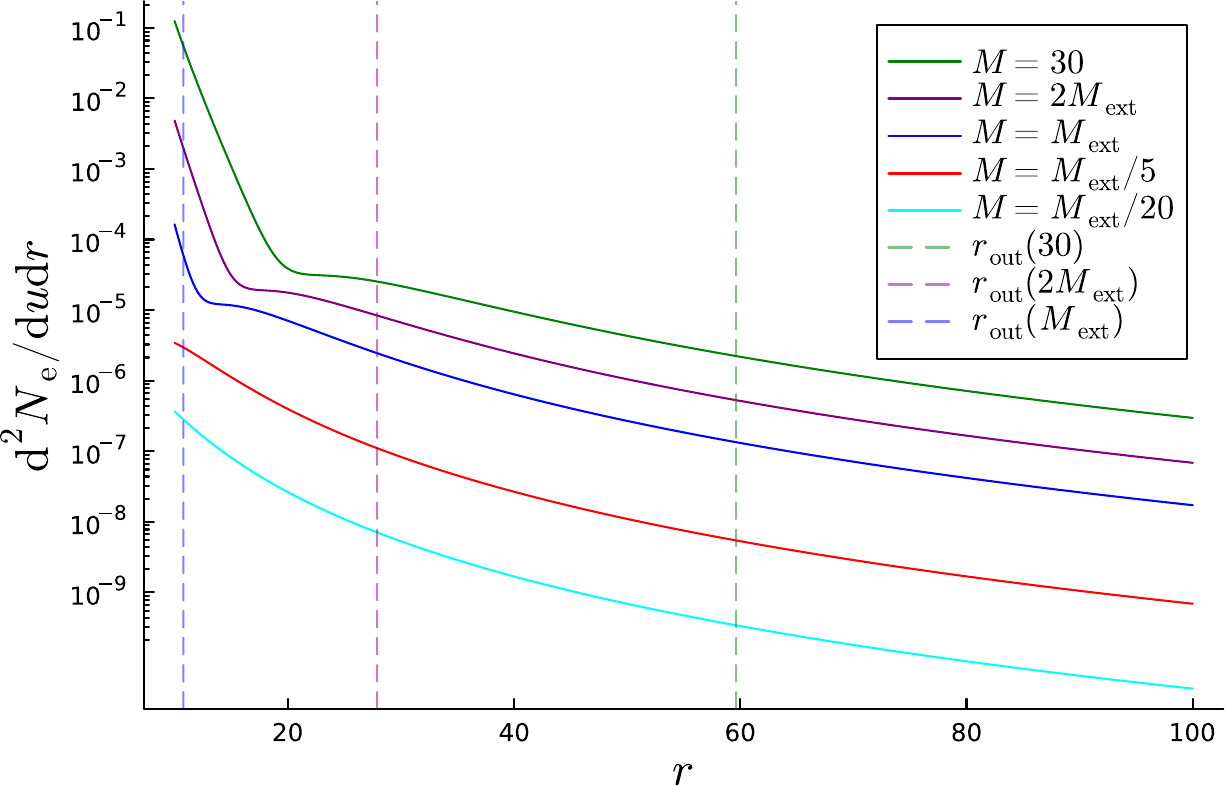}
    \caption{{Radial profiles of the pair-production event rate from Eq. \eqref{dNdudr}, evaluated in the quantum-corrected black hole geometry $f_\mathrm{qS}(r)$ for several mass values: two-horizon black holes ($M > M_\mathrm{ext}$), an extremal black hole ($M = M_\mathrm{ext}$), and horizonless configurations ($M < M_\mathrm{ext}$). In all cases, the rate grows for smaller $r$, where spacetime curvature is stronger. For configurations with horizons, the vertical dashed lines mark the locations of the outer horizons, $r_\mathrm{out}(M)$. The plot is generated in Planck units with $\gamma = 1$.}}
    \label{fig_dNdudr}
\end{figure}
{In Fig. \ref{fig_dNdudr}, we plot the emission rate \eqref{dNdudr} for the quantum-corrected black hole model analyzed in the following sections. Notably, this plot demonstrates that pair production persists even when horizons are absent from the spacetime.}

The luminosity (energy emission rate) is 
\begin{equation} \label{flux2}
 \frac{\partial E}{\partial u} =    2 \int_{r_*}^{\infty} \mathrm{d}\tilde{r}   \frac{\mathrm{d}^2 N_{\mathrm{e}}}{ \mathrm{d}u \,  \mathrm{d}\tilde{r}} f_\mathrm{ppe} f_\mathrm{esc} E_\mathrm{curv} \,,
\end{equation}
where the factor of $2$ introduced in Eq. \eqref{flux2} converts the pair-creation rate into the particle-production rate. The factor $f_{\mathrm{ppe}} = 12/\pi^2$ accounts for the ratio between the rate density of particle pairs and the rate density of production events, as derived in \cite{Wondrak:2023zdi}. {In the integral in Eq. \eqref{flux2}, the lower limit $r_*$ must be chosen very close to the outer horizon $r_\mathrm{out}$. Within this static-background approximation, particles produced inside the horizon ($r < r_\mathrm{out}$) are reabsorbed by the black hole and therefore do not contribute to the outgoing flux.} We further assume that for particles {produced at $r>r_*$}
\begin{enumerate}[label=(\roman*)]
\item The escape probability is given by $f_\mathrm{esc} =1/2$, corresponding to purely radial particle emission where half of the particles are emitted outward and escape the black hole, while the other half are emitted inward.
\item For simplicity, we neglect the gravitational redshift experienced by the particles as they propagate from the emission point to infinity.
\end{enumerate}
For a particle produced, we assign a characteristic energy scale $E_\mathrm{curv}$ determined by comparing the imaginary part of the effective Lagrangian with the corresponding expression for the electromagnetic Schwinger effect (see the discussion around Eq. (10) in \cite{Wondrak:2023zdi}\footnote{Note also that because we focus on non-Ricci-flat spacetimes, $E_\mathrm{curv}$ takes a more general form than the expression given in Eq. (10) of Ref. \cite{Wondrak:2023zdi}.})
\begin{equation}
E_\mathrm{curv} = \hbar \left[ \frac{ 12 \operatorname{Im}\left(  \mathcal{L}_\mathrm{} \right)/\left(  \frac{64 \pi}{\hbar}\right)}{2} \right]^{\frac{1}{4}} \,. 
\end{equation}
Having completed the preceding steps, the mass-loss rate $\frac{\partial M }{\partial u} = -\frac{\partial E }{\partial u}$ due to {Hawking-like radiation} for the Schwarzschild-like geometry $f_\mathrm{qS}(r)$ is given by
\begin{equation} \label{dMdu_schwinger}
\begin{aligned}
\frac{\partial M}{\partial u} = -\int_{r_*}^\infty \dd r \frac{2^{7/4} \hbar (G M)^{5/2}}{5^{5/4} \pi^2 r^{13}} \Big(&r^6 - 20 G M r^3 \gamma^2 \Delta \\
&+ 126 G^2 M^2 \gamma^4 \Delta^2\Big)^{5/4} \,,
\end{aligned}
\end{equation}
where the integral accounts for the energy emission rate carried by particles produced in the radial range from $r_*$ to infinity. Thus, the Schwinger approach provides an alternative framework for $M(u)$ estimation. Interestingly, pair production occurs as long as the curvature invariants in the effective action \eqref{effective_action} are non-zero, or, more physically, as long as tidal forces are present in the spacetime under consideration. This implies that the presence of an event or even a trapping horizon is not a necessary condition for particle production\footnote{This insight motivated subsequent work \cite{Falcke:2024rdu}, building upon \cite{Wondrak:2023zdi}, to estimate the lifetimes of horizonless astrophysical objects.}.

We emphasize that the escape factor $f_\mathrm{esc}$ adopted in the present work differs from that considered in Ref. \cite{Wondrak:2023zdi} in their analysis of gravitational particle production in a Schwarzschild spacetime. In their approach, particle emission is assumed to be isotropic in the local rest frame of a static Schwarzschild observer, and the escape probability is determined by the solid angle of the capture cone, resulting in a radius-dependent escape factor. In the present work, by contrast, we model the backreaction through spherically symmetric radial null dust. This effective description naturally motivates restricting the emission to radial directions, corresponding to an $s$-wave approximation of quantum field propagation. To qualitatively compare the resulting luminosity estimates, we evaluate the luminosity obtained from Eq.~\eqref{dMdu_schwinger} in the classical limit $\Delta\to 0$ and $r_*\to 2GM$, yielding $\frac{\partial E}{ \partial u} \approx 44.8 \cdot10^{-5} \; \hbar (GM)^{-2}$ (compare with Eq. (11) in \cite{Wondrak:2023zdi}).

\begin{figure}
    \centering
    \includegraphics[width=1\linewidth]{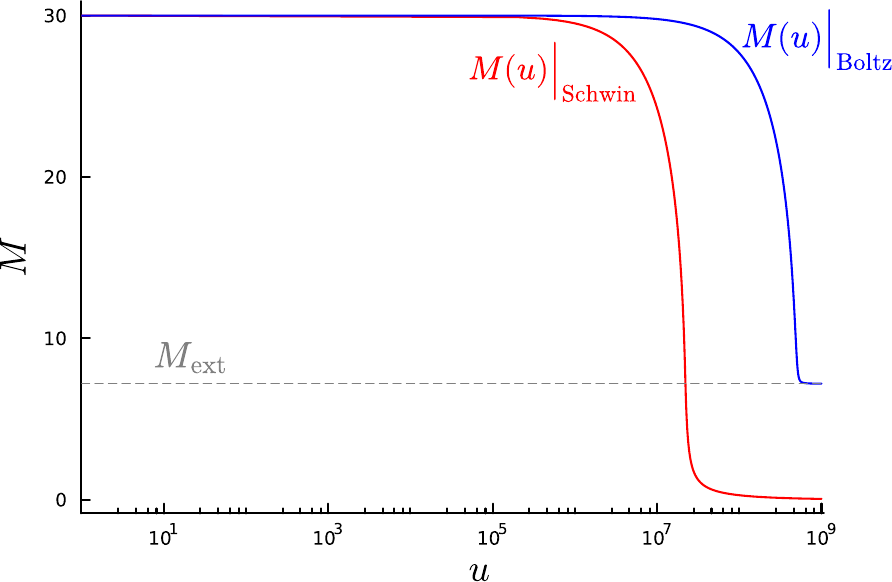}
    \caption{{Evaporation history of the mass parameter $M(u)$ as a function of the $u$-time parameter for a quantum-corrected $f_{\mathrm{qS}}$ black hole, comparing two evaporation laws within a quasi-static approximation. The blue curve, $M(u)\big\vert{}_\mathrm{Boltz}$, is obtained using the Stefan–Boltzmann evaporation law, where the mass decreases and asymptotically approaches the extremal black hole mass $M_\mathrm{ext}$ (dashed gray line). In contrast, $M(u)\big\vert_{\mathrm{Schwin}}$ corresponds to the Schwinger evaporation approach. At the furthest plotted time, the mass remains nonzero, though very close to zero. The plot is generated in Planck units with an initial mass $M(0)=30$ in both cases and $\gamma=1$. } }
    \label{fig_MM_comparison}
\end{figure}

In the approach of \cite{Wondrak:2023zdi, Falcke:2024rdu}, gravitational pair production generates particles whose trajectories determine whether they escape to infinity or are captured and absorbed by the compact object. Unlike the conventional Hawking mechanism—which is characterized by an ingoing flux of negative energy—this absorbed component represents a positive-energy deposition into the object. In the present work, we neglect the backreaction of these absorbed positive-energy particles. However, the authors argue that energy conservation in gravitational pair creation requires negative-energy modes, potentially associated with violations of the classical energy conditions similar to those in the Casimir effect---see the discussion section of \cite{Falcke:2024rdu}. These modes (negative-energy partners) supply the inward negative-energy flux toward the compact object while positive-energy modes form the outgoing radiation, and it is precisely this negative-energy flux that we employ in Sec. \ref{sec_hawking_israel}.

{We emphasize that the mass functions $M(u)$ plotted in Fig.~\ref{fig_MM_comparison} are not necessarily realized across their entire domain in the complete picture of black hole formation and evaporation presented in Sec.~\ref{sec_completespacetime}. In the Schwinger case, for instance, evaporation terminates well before the maximal $u$-time shown in Fig.~\ref{fig_MM_comparison}.}

\begin{figure}
    \centering
 \includegraphics[width=1\linewidth]{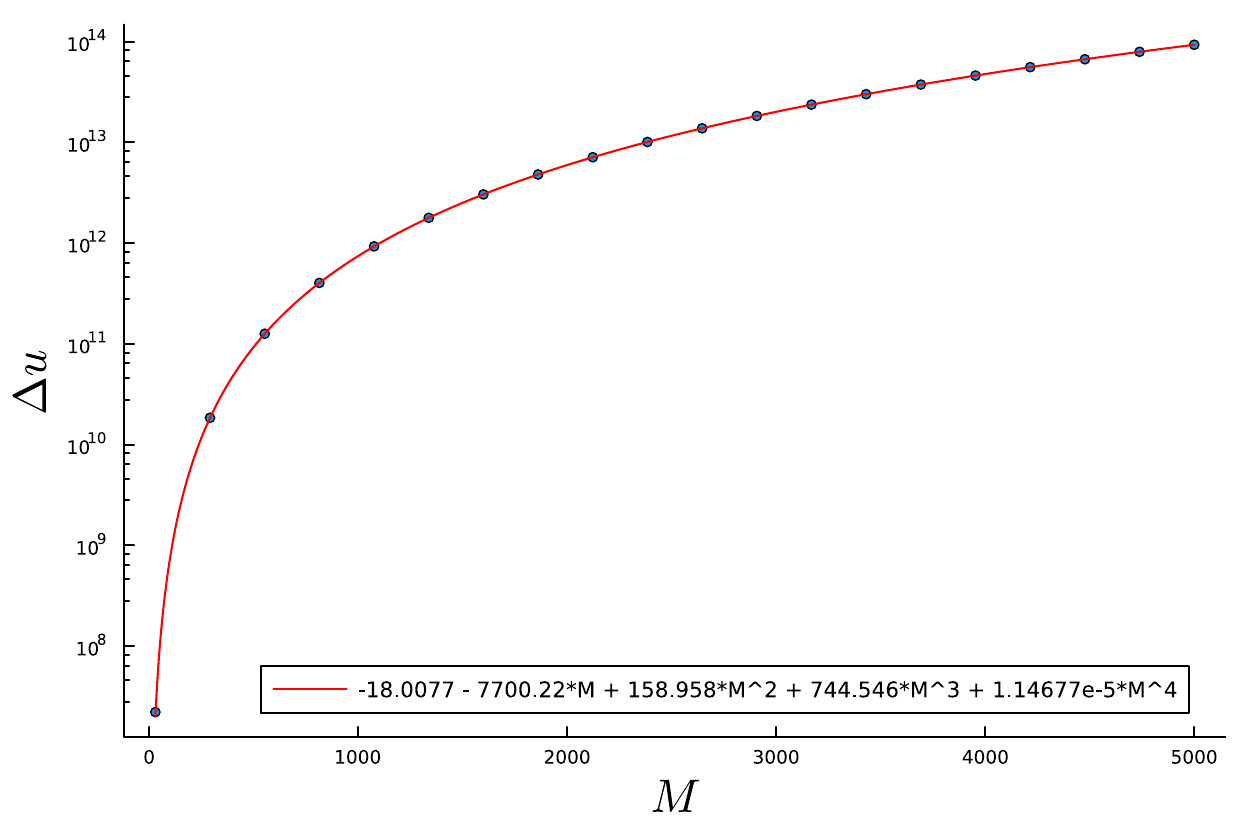}
    \caption{Evaporation times $\Delta u$ as a function of the black hole mass $M$ (blue circles) for the $f_\mathrm{qS}$ black hole geometry, computed using the Schwinger evaporation law, alongside a polynomial fit (red curve). Results are shown in Planck units for $\gamma = 1$.}
    \label{fig_fit}
\end{figure}

{The Schwinger evaporation law also allows us to estimate the lifetime of higher-mass black holes—or more precisely, the lifetime of the dynamical trapping region, since no event horizons are present in the complete spacetime picture as it is discussed in Sec. \ref{sec_completespacetime}. In Fig. \ref{fig_fit}, we plot these lifetimes, defined as the finite $u$-time required for the mass to decrease to the extremal value $M_\mathrm{ext}$, at which point the horizons become degenerate and subsequently disappear.}

\section{Hawking-like evaporation in the Darmois-Israel formulation} \label{sec_hawking_israel}

In Sec. \ref{sec_evaporationlaws} we have derived two possible evaporation laws for the black hole exterior, i.e. the backreaction of the outgoing positive flux of Hawking-like particles on the geometry. In this section we relate such results to the backreaction of the ingoing negative flux of Hawking-like partners to the geometry inside the outer {trapping} horizon. The aim for this is twofold: first, having a complete picture of the spacetime outside the collapsing thin shell. Then, the dynamical spacetime inside the outer horizon will be necessary to derive the thin shell motion from the Darmois-Israel junction conditions. Out of the general class of ingoing \eqref{vaidya} and outgoing \eqref{vaidyaout} Vaidya effective models introduced in Sec. \ref{sec_nulldust}, we specialize the discussion to the effective model corresponding to $f_\mathrm{qS}(r)$ Schwarzchild-like geometry discussed in Sec. \ref{sec_lqgdustcollapse}. By performing the Lemos' trick on the specific model given by the shift vector \eqref{shift}, we get
\begin{equation}\label{invad}
\begin{aligned}
  \dd s^2 =&-\left({1-\frac{2GM(v)}{r}+\frac{\gamma^2 \Delta (2GM(v))^2}{r^4} } \right)\dd v^2  \\ 
  &+2 \dd r \dd v +r^2 \dd \Omega^2 \,,  
 \end{aligned}
 \end{equation}
 \begin{equation}
 \begin{aligned}
  \dd s^2 =&-\left({1-\frac{2GM(u)}{r}+\frac{\gamma^2 \Delta (2GM(u))^2}{r^4} } \right)\dd u^2  \\ 
  &-2 \dd r \dd u +r^2 \dd \Omega^2
   \label{outvad} \,,
\end{aligned}
\end{equation}
respectively for the ingoing and outgoing effective Vaidya solutions. 

\begin{figure}
    \centering
    \includegraphics[width=0.85\linewidth]{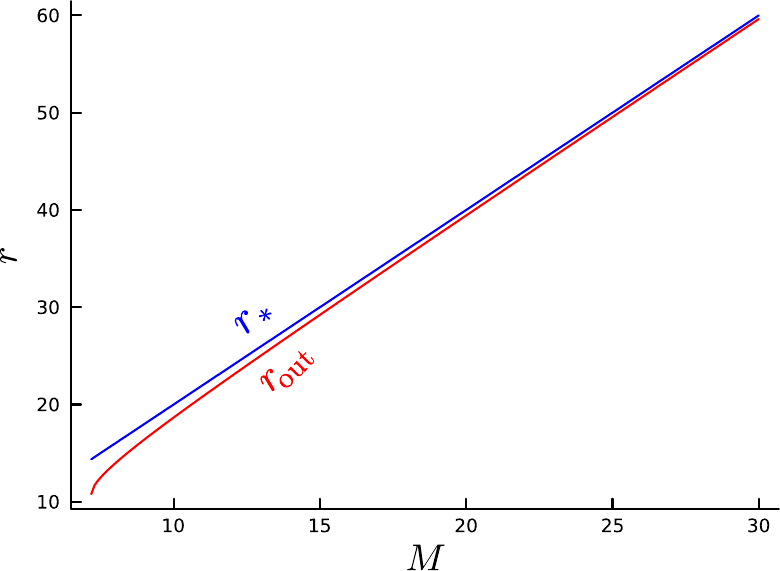}
    \caption{The functions $r_*(M)$ and $r_\mathrm{out}(M)$—representing the areal radius on the junction surface $\Sigma_+$ and the position of the outer trapping horizon, respectively—are plotted for $M > M_\mathrm{ext}$. We set $\gamma = 1$ and employ Planck units.}
    \label{fig_horizon_location}
\end{figure}

To get a consistent spacetime across the evaporation surface, we will impose the continuity of the induced metric on a dynamical boundary between the
outgoing Vaidya solution \eqref{outvad} (that we will denote with $\mathcal{M}_1$) and the ingoing Vaidya \eqref{invad} (that we will denote with $\mathcal{M}_2$). {As the boundary separating the two spacetimes, we choose the dynamical hypersurface $\Sigma_+$ given by {following areal radii}
\begin{equation}
r_*\Big|_\mathrm{Boltz}=2GM(v)=2GM(u) \,,
\end{equation}
and 
\begin{equation}
r_*\Big|_\mathrm{Schwin}= \begin{cases} 2GM\; \text{for}\; M= M(u)= M(v) \geq M_\mathrm{ext} \,, \\
2GM_\mathrm{ext}\; \text{for}\;  M(u)= M(v) < M_\mathrm{ext} \,,
\end{cases}
\end{equation}
where ``Boltz" and ``Schwin" denote the surface choices for the Stefan-Boltzmann and Schwinger evaporation laws described in Sec. \ref{sec_evaporationlaws}, respectively. The difference lies in the fact that evaporation terminates at $M_\mathrm{ext}$ in the former approach, whereas it continues beyond that point in the latter. However, once trapping horizons disappear, the exact location of the evaporation surface becomes arbitrary\footnote{Here, the specific constant radius $2GM_\mathrm{ext}$ is chosen to ensure surface continuity and preserve its timelike character.}.} Note that such a boundary, {in the both cases,} does not describe the location of the effective {trapping} horizon, which instead, for macroscopic black holes $2GM(v)\gg\sqrt{\Delta}$ is located at
\begin{equation}
r_\mathrm{out}(v)=2GM(v)-\frac{\gamma^2 \Delta}{2GM(v)}+ O\left[\frac{\Delta^{{4}/{3}}}{(2GM(v))^{5/3}} \right] ~,   
\end{equation}
as one can easily check by computing $g_{vv}=0$ in \eqref{vaidya}, so slightly inside $2GM(v)$ (if one computes it using the effective outgoing Vaidya metric gets the same result), and for any mass $M(v)$ holds $r_\mathrm{out}<r_*$, signalling that during the evaporation, $\Sigma_+$ lies always outside (even though slightly) the outer horizon.
In general, it is not said that the mass function remains continuous across the boundary $r_*$ during the evolution of the two Vaidya spacetimes. The continuity of the mass function is instead compatible with the absence of surface density on $\Sigma_+$, as verified through the second Darmois-Israel junction condition. We will come back to this point in Sec. \ref{sec_completespacetime}.
 As a side note, the effective metric actually retains another horizon, an inner one, located, for macroscopic black holes, at
\begin{equation}
r_\mathrm{in}(v)=(\gamma^2 \Delta 2G M(v))^\frac{1}{3}+O\left[\frac{\Delta^{2/3}}{(2GM(v))^{1/3}} \right]\,,    
\end{equation}
but whether and when this inner horizon is developed during the spacetime evolution depends on the dynamics of the thin shell, and we will deal with this in Sec. \ref{sec_completespacetime}. Before moving on, let's justify the location of $r_*$ from physical grounds. Keeping in mind that we are modeling Hawking pair production near the dynamical horizon, it is reasonable to plug the pair production surface just outside the {trapping} horizon. Hawking-like partners, which carry negative energy (modeled by collapsing Vaidya with negative energy density), violating the null energy condition, will propagate in a non-trapped region (in $r_\mathrm{out}(v)<r<r_*(v))$, which, however, can be easily shown to be extremely narrow, for any mass $M(v)$ during the evaporation. Fig. \ref{fig_horizon_location} shows the location of $r_*$ and $r_\mathrm{out}$ for any possible black hole mass, proving that the assumption on the location of $\Sigma_+$ is physically justified.

Then we impose continuity of the induced metric on such a boundary (the first Darmois-Israel junction condition). The boundary, as seen by the two metrics, is parametrized by: $\{u_{(1)}=U_{(1)},r_{(1)}=r_*(U_{(1)}), \theta_{(1)}, \phi_{(1)} \}$, $\{V_{(2)}=V_{(2)}, r_{(2)}=r_*(V_{(2)}), \theta_{(2)}, \phi_{(2)} \}$. Therefore, the pullback of the two metrics on the boundary is provided by: $\dd r_{(1)}=2G\frac{\dd M(U_{(1)})}{\dd U_{(1)}} \dd U_{(1)}$, $\dd r_{(2)}=2G\frac{\dd M(V_{(2)})}{\dd V_{(2)}} \dd V_{(2)}$. Substituting in the metrics, we get
\begin{equation}
\begin{aligned}
  \dd s_{(2)}^2\bigg|_{\Sigma_+} = -\left[\frac{\gamma^2 \Delta}{(2GM(V_{(2)}))^2} -  4G\frac{\dd M(V_{(2)})}{\dd V_{(2)}} \right]\dd V_{(2)}^2 \\  + \left[2G M(V_{(2)})\right]^2\dd \Omega^2_{(2)}  \,,
 \end{aligned}
 \end{equation}
 \begin{equation}
 \begin{aligned}
  \dd s_{(1)}^2\bigg|_{\Sigma_+} =-\left[\frac{\gamma^2 \Delta}{ (2GM(U_{(1)}))^2} +4G\frac{\dd M(U_{(1)})}{\dd U_{(1)}} \right]\dd U_{(1)}^2 \\ + \left[2GM(U_{(1)})\right]^2 \dd \Omega_{(1)}^2 \,.
\end{aligned}
 \end{equation}
Then we require continuity, which gives in the angular part $\dd \Omega_{(1)}=\dd \Omega_{(2)}$. The condition $2GM(U_{(1)})=2GM(V_{(2)})$ can be differentiated, giving
\begin{equation}
  \frac{\dd M}{\dd V_{(2)}}= \frac{\dd U_{(1)}}{\dd V_{(2)}}\frac{\dd M}{\dd U_{(1)}} ~,
\label{reluv}
\end{equation}
and the first Darmois-Israel condition reads
\begin{align}
&\frac{\gamma^2 \Delta}{ (2GM(U_{(1)}))^2} -4G\frac{\dd M(U_{(1)})}{\dd U_{(1)}}\frac{\dd U_{(1)}}{\dd V_{(2)}} = \notag\\
&=\left(\frac{\dd U_{(1)}}{\dd V_{(2)}}\right)^2 \left[ \frac{\gamma^2 \Delta}{ (2GM(U_{(1)}))^2} +4G\frac{\dd M(U_{(1)})}{\dd U_{(1)}}  \right]\,.
\label{ciao}
\end{align}
Once $M(U_{(1)})$ is fixed as a solution of the specific evaporation law $\dd M(U_{(1)})/ \dd U_{(1)}$ chosen, the previous is a second degree equation for $\dd U_{(1)}/\dd V_{(2)}$. It has in general two solutions: the first, $dU_{(1)}/dV_{(2)}=-1$, is not physically valid, since would require a gluing where one of the two fluxes is considered time reversed. The other solution is
\begin{equation}
  \frac{\dd U_{(1)}}{\dd V_{(2)}}= \frac{\Delta \gamma^2}{\Delta \gamma^2 + 16 G^3 M(U_{(1)})^2 \frac{\dd M(U_{(1)})}{\dd U_{(1)}}}  ~.
 \label{jacob}
\end{equation}
This determines completely $M(V_{(2)})$, so the geometry of $\mathcal{M}_2$---the exterior of the thin shell---and can be used to study the thin shell evolution. To give a concrete idea of the rate of mass evaporation between $\mathcal{M}_1$ and $\mathcal{M}_2$, {we can write explicitly \eqref{jacob} for macroscopic black holes, for the Stefan-Boltzmann case
\begin{equation} \label{FFFBoltz}
\begin{aligned}
 { \frac{\dd U_{(1)}}{\dd V_{(2)}}\Bigg|_\mathrm{Boltz}} =   \frac{11520 \pi^2 \gamma^3}{  11520 \pi^2 \gamma^3 - \sqrt{3}}  + O\left[\frac{\sqrt{\Delta}}{2GM(U_{(1)})} \right]^2 \,.
\end{aligned}
\end{equation}

In the vacuum case, the match can be easily performed just outside the horizon, since the asymptotic region is covered by both the charts, without the introduction of any thin shell. 
In our case, the situation is more subtle, since the dynamical nature of $\mathcal{M}_1$ and $\mathcal{M}_2$, specifically because the two regions are not physically equivalent (one is filled by negative null dust, one by positive null dust). For this reason, a surface stress layer is needed in our case, while in the vacuum is absent (a similar result is found in \cite{Hayward:2005gi}, even though the underlying effective model is different). We will come back to this point in Sec. \ref{sec_completespacetime}. In conclusion, the specific choice of $r_*$ adopted here does not restrict the generality of the framework, as selecting a different (dynamical) hypersurface just outside the dynamical outer horizon leaves the results qualitatively unchanged.
With this in hand, we can now turn the analysis to the thin shell evolution.
\section{Dynamics of the evaporating thin shell}
\label{secdynamicsthin}
 In Sec.\ref{sec_hawking_israel}, we modeled {Hawking-like evaporation} by imposing the first Darmois-Israel junction condition across a boundary lying slightly outside the evolving trapping horizon, using an effective ingoing Vaidya metric to model the flux of Hawking-like partners and an effective outgoing Vaidya metric for Hawking-like particles. However, since our aim is to obtain a complete picture of matter collapse, black hole formation, and {Hawking-like evaporation}, we must also include the matter collapse phase of the spacetime.

As stated in the introduction, we model stellar collapse using a timelike dust thin shell. This shell forms the trapping horizon during its collapse, and {Hawking-like evaporation} commences immediately afterward. The dynamics of the thin shell is derived via the Darmois-Israel formulation applied to the shell's surface, which separates two distinct spacetimes that we denote as $\mathcal{M}_3$ and $\mathcal{M}_2$. Specifically, $\mathcal{M}_2$ is the same submanifold considered in the previous section---namely, a collapsing Vaidya spacetime composed of null dust with negative energy density to model the flux of Hawking-like partners---while $\mathcal{M}_3$ is Minkowski spacetime. Because Hawking-like partners carry negative energy density, their production decreases the gravitational mass of the black hole (or equivalently the thin-shell gravitational mass), and also the inertial mass of the thin shell. This effect is accounted for by covariant energy-momentum conservation across the thin shell.

To derive the thin shell dynamics, we employ the Hamiltonian formulation of the Darmois-Israel junction conditions, initially introduced at the effective level in \cite{Fazzini:2025nse} for an isolated thin shell. To simplify the analysis, we assume that the thin shell starts from the trapping horizon at the initial time, thereby eliminating the need to alter the exterior spacetime during its evolution.

It is important to note that because the Hamiltonian formulation of the Darmois-Israel junction conditions---specifically the second condition---requires a timelike foliation, we cannot work directly with the exterior Vaidya metric. Following Lemos' approach applied to the junction conditions, we first consider for the exterior an LTB expanding dust model in the non-marginally bound case with a generic constant parameter $\varepsilon > 0$. Once the equations are derived, {we} apply Lemos' trick.

\subsection{First junction condition for the thin shell}
 The first Darmois-Israel junction condition is straightforward to derive. The approach requires the continuity of the junction 3-metric. In principle, the boundary can be assumed to be a timelike, null, or spacelike hypersurface; however, since we require a timelike dust thin shell situated on this boundary, we choose the first option. To this end, we consider the induced metric on this 3-dimensional timelike hypersurface (which we denote by $\Sigma_-$ to distinguish it from the junction surface near the horizon analyzed previously)
\begin{equation}
 \dd s^2\bigg|_{\Sigma_{-}}=-\dd \tau^2+ R(\tau)^2 \dd \Omega^2 ~,  
\end{equation}
where $R(\tau)$ denotes the areal radius of the shell. 

Note that specifying this form for the induced metric automatically enforces the timelike requirement (had it been a null surface, the metric would be 2-dimensional, whereas a spacelike surface would possess a Euclidean signature $(+++)$). From a physical perspective, $\tau$ can be identified as the proper time of the dust shell, since for $\theta = \text{const.}$ and $\phi = \text{const.}$ (i.e., an observer comoving with the evolving surface), variations in $\tau$ coincide with the invariant interval $\mathrm{d}s$. To impose continuity across $\Sigma_-$, we perform the pullback of the interior and exterior metrics onto $\Sigma_-$ via the embedding relations $\{t_{(3)} = T_{(3)}(\tau), r_{(3)} = R_{(3)}(\tau), \theta_3, \phi_3\}$ and $\{T_{(2)} = T_{(2)}(\tau), r_{(2)} = R_{(2)}(\tau), \theta_{(2)}, \phi_{(2)}\}$, respectively, where the subscript $2$ refers to the exterior ingoing Vaidya metric and the subscript $3$ to the interior Minkowski metric.{ The Minkowski metric will be written also in PG-like gauge \eqref{minkopg}, for a reason that will be clear later.} We obtain
\begin{widetext}
\begin{equation}
\begin{aligned}
  \dd s_{(2,3)}^2=- \left[ \left(1-\frac{(N^r_{(2,3)})^2}{1+\varepsilon} \right)\Dot{T}_{(2,3)}^2 -\frac{2}{1+\varepsilon}N^r_{(2,3)} \Dot{T}_{(2,3)} \Dot{R}_{(2,3)} -\frac{\Dot{R}^2_{(2,3)}}{1+\varepsilon}   \right]\dd \tau^2 + R_{(2,3)}(\tau)^2 \dd \Omega_{(2,3)}^2 ~,
\end{aligned}
\end{equation}
\end{widetext}
{where $\varepsilon$ is the same in both sides. Note that for the exterior, $\varepsilon$ is a physical quantity related to the kinetic energy of the shells at $r=+\infty$, while for the interior is a gauge degree of freedom.} $N^r_{(3)}=0$, but we keep it for notational convenience.

As anticipated, we first consider the PG-like metric for non-marginally bound collapsing timelike dust as the exterior spacetime. Once the equations are derived, we perform Lemos' trick to get $\mathcal{M}_2$ being filled by null dust. Requiring the continuity of the metric across the surface $\Sigma_-$ then leads to the following:
\begin{equation}
\begin{cases}
& R_{(2)}(\tau)=R_{(3)}(\tau)=R(\tau) ~,\\
& \theta_{(2)}=\theta_{(3)}=\theta~,\\
&\phi_{(2)}=\phi_{(3)}=\phi~, \\
&\left(1-\frac{(N^r_{(2,3)})^2}{1+\varepsilon} \right)\dot{T}_{(2,3)}^2-\frac{2}{1+\varepsilon}N^r_{(2,3)} \dot{T}_{(2,3)} \dot{R}_{(2,3)} -\frac{\dot{R}^2_{(2,3)}}{1+\varepsilon} =1 \,.
\end{cases}
\label{sistema}
\end{equation}
The previous system consists in the first Darmois-Israel junction condition imposed on a time-like boundary.

We can define the shell $4$-velocity as
\begin{equation}
 u^\mu_{(2,3)}= \left(\frac{\dd T_{(2,3)}}{\dd \tau}, ~ \frac{\dd R}{\dd \tau},~0,~0 \right)  ~. 
\end{equation}
Then, it is immediate to see that the last condition in \eqref{sistema} is simply the normalization of the (timelike) shell $4$-velocity.

\subsection{Second junction condition for the thin shell}
 Effective models inspired by loop quantum gravity are usually developed, as is the fundamental theory, within the canonical formulation. Consequently, following the standard path of the Darmois-Israel formulation \cite{Poisson:2009pwt} would require first recasting the equations in a 4-dimensional formulation and then imposing the second junction condition. However, it can be shown both in the classical case \cite{Crisostomo:2003xz} and in the effective case \cite{Fazzini:2025nse} that an alternative formulation of the second Darmois-Israel junction condition exists within the canonical framework.

The core idea is to solve the constraints of the system in integral form, such that potential discontinuities and singular surfaces in the constraints are treated distributionally, in the same spirit as the standard formulation of the second junction condition
\begin{equation}
\begin{aligned}
 &  \int_{{R}-\epsilon}^{{R}+\epsilon} N \left(\mathcal{H}^\mathrm{grav}+ \mathcal{H}^\mathrm{matt}+ \mathcal{H}^\mathrm{shell} \right) \dd r \approx 0 \,,  \\
 & \int_{{R}-\epsilon}^{{R}+\epsilon} N^r \left(\mathcal{D}^\mathrm{grav}+ \mathcal{D}^\mathrm{matt}+ \mathcal{D}^\mathrm{shell} \right) \dd r \approx 0  ~,  
\end{aligned}
\end{equation}
where $\epsilon$ is a small radial distance, $R$ is the dynamical shell location, and $\approx$ denotes equality in the Dirac weak sense, meaning it holds only on-shell. Here, the constraints receive three contributions: the gravitational contribution, the matter contribution from the dust-filled region (which is assumed to retain its classical functional form even at the effective level, in line with effective models in LQG), and the thin shell contribution (which also preserves its classical functional form). For notational convenience, we write
\begin{equation}
\begin{aligned}
   \mathcal{H}:= & ~\mathcal{H}^\mathrm{grav}+ \mathcal{H}^\mathrm{matt}+ \mathcal{H}^\mathrm{shell} ~, \\
  \mathcal{D} :=& ~\mathcal{D}^\mathrm{grav}+ \mathcal{D}^\mathrm{matt}+ \mathcal{D}^\mathrm{shell}~.
\end{aligned}
\end{equation}
The constraints are typically solved after spatial integration. However, if the spatial gauge is the same across the boundary (as in our case), one can first solve the diffeomorphism constraint in differential form and rewrite the spatially gauge-fixed Hamiltonian constraint \cite{Fazzini:2025nse,Bobula:2026zlq}. This is the approach we adopt here. We begin with the classical theory (introduced in Sec.~\ref{sec_lqgdustcollapse}) under the areal gauge described in Sec.~\ref{sub_sec_arealgauge}, and subsequently extend it to the effective model.
The constraints are given by \cite{Bobula:2026zlq}
\begin{align}
&\mathcal{H}^\mathrm{grav}={-\frac{1}{\sqrt{1+\varepsilon}}\partial_r \tilde{H}_\mathrm{GR}}+ N^r( \mathcal{D}^\mathrm{shell}+\mathcal{D}^\mathrm{matt} ) \,,\\
  &\mathcal{H}^\mathrm{matt}= { 4 \pi  p_T \theta(r-R) }\,, \\
 & \mathcal{D}^\mathrm{matt}=0 \,,
\end{align}
where $\tilde{H}_\mathrm{GR}$ in PG-like coordinates retains the same expression as in the LTB gauge \eqref{Hclassical}, whereas the matter contribution from $\mathcal{M}_2$ is evaluated after imposing the dust-time gauge $T - t = 0$, with the shell contribution to the constraints specified subsequently. The dust-time gauge in the exterior requires $N = 1$, and the same value of the lapse is necessary to recover the PG-like for of the Minkowski metric in the interior.

We can therefore write the resulting Hamiltonian constraint \begin{align}
 \mathcal{H}=&{-\frac{1}{\sqrt{1+\varepsilon}}\partial_r \tilde{H}_\mathrm{GR}} + N^r \mathcal{D}^\mathrm{shell}  \notag\\
&+\mathcal{H}^\mathrm{shell}+4 \pi p_T \theta(r-R)~,
\end{align}
where the first term applies to both the interior and exterior regions, although it vanishes identically in the interior. Upon taking the limit $\epsilon \to 0$ after integration, only terms containing spatial derivatives and Dirac delta functions survive, in particular, the last term vanishes. Let us now examine the explicit form of the shell contributions to the constraints.
To construct them, we begin with the energy-momentum tensor of a dust thin shell, which reads
\begin{equation}
 T_{\mu \nu}=\sigma u_\mu u_\nu \delta(\chi) ~,
\end{equation}
where $\sigma$ is the proper surface density of the shell and $u^\mu$ is the shell $4$-velocity. Requiring $\sigma$ to be the proper surface density entails evaluating it in the shell's rest frame, in which the spatial radial coordinate orthogonal to the shell's spatial surface is denoted by $\chi$. In this coordinate system, the shell is located at $\chi = 0$ at all times. The scalar and diffeomorphism contributions to the constraints can be readily computed by projecting the energy-momentum tensor along the $tt$ and $tr$ directions, respectively. This first requires defining the 4-velocity adapted to the foliation (which corresponds to the PG-like foliation in the exterior and the Minkowski foliation in the interior). This velocity is easily obtained by computing the normal to the $t = \text{const.}$ surfaces: $n_\mu = -\partial_\mu t = (-1, 0, 0, 0)$, where the minus sign serves to align the covector with the direction of the time flow.

The corresponding four-vectors are defined as $n^\mu_{(2,3)} = \left(1,-N^r_{(2,3)}, 0, 0\right).$
Although these four-vectors serve as local four-velocities for timelike observers in their respective spacetimes, they generally differ from $u^\mu$, the shell's four-velocity, because the thin shell does not, in general, follow geodesic trajectories---a behavior that directly mirrors the classical regime.

At this stage, it becomes evident that the contribution of the thin shell to the constraint equations depends on the side of the boundary $R$ upon which it is evaluated. This dependence stems from the non-smoothness of the four-dimensional spacetime across the shell, induced by the jump in the extrinsic curvature. Furthermore, the interior and exterior coordinate times generally do not coincide at the boundary, in agreement with the first Darmois-Israel junction condition. This mismatch, however, does not hinder the integral formulation of the constraints, as the integration is performed strictly along $r$---a coordinate guaranteed to be continuous by the first junction condition within these coordinate charts. Equipped with these results, we now compute the scalar contribution of the thin shell to the constraints
\begin{align}
(\mathcal{H}^\mathrm{shell})_{(2,3)}&=4 \pi\sigma \sqrt{q}_{(2,3)} n^\mu_{(2,3)} n^\nu_{(2,3)} u^{(2,3)}_\mu u^{(2,3)}_\nu  \delta(\chi_{(2,3)})  \notag \\
 &=4\pi \sigma \sqrt{q}_{(2,3)} (n \cdot u)^2_{(2,3)}  \delta(\chi_{(2,3)}) \,,
\end{align}
where $\sqrt{q}_{(2,3)} = r^2/\sqrt{1+\varepsilon}$ denotes the square root of the spatial metric determinant, which takes the same form for both the interior and exterior metrics. Writing the Minkowski metric in PG-like coordinates ensures the absence of jumps in the three-metric determinant across the shell. Furthermore, the coordinate $\chi$ is defined differently on either side of the boundary because the associated four-vector is orthogonal to the thin-shell surface, which is itself orthogonal to the shell's four-velocity—the latter differing depending on whether it is evaluated using the interior or exterior metric. Analogously, one can derive the diffeomorphism contribution of the shell to the constraints \cite{Fazzini:2025nse, Bobula:2026zlq}
\begin{equation}
(\mathcal{D}^\mathrm{shell})_\pm=4\pi \sqrt{q}_{(2,3)} (n\cdot u)_{(2,3)} u_r^{(2,3)} \sigma \delta(\chi_{(2,3)}) ~. 
\end{equation}
With these results in hand, we can write the explicit form of the classical Hamiltonian constraint
\begin{align}
&\mathcal{H}=  {-    \frac{1}{\sqrt{1+\varepsilon}}\partial_r \tilde{H}_\mathrm{GR} }  
+\notag \\
&+4 \pi \frac{r^2}{\sqrt{1+\varepsilon}}\sigma \delta(\chi_{(2,3)})\left[N^r_{(2,3)}(n\cdot u)_{(2,3)} u_r^{(2,3)} + (n\cdot u)_{(2,3)}^2 \right] ~.
\end{align}
Everything done so far pertains to the classical setup. To obtain the effective solution, we need to polymerize the gravitational part of the Hamiltonian and the shift vector itself. Specifically, we perform the substitution $\tilde{H}_\mathrm{GR} \to \tilde{H}_\mathrm{eff}$ (see Sec.~\ref{sec_lqgdustcollapse}) and apply an analogous replacement for $N^r$. The effective Hamiltonian constraint then reads
\begin{widetext}
\begin{equation}
\begin{aligned}
  \mathcal{H}= {  - \frac{\partial_r \tilde{H}_\mathrm{eff}}{\sqrt{1+\varepsilon}}} 
  +4 \pi \frac{r^2}{\sqrt{1+\varepsilon}}\sigma \delta(\chi_{(2,3)})\left[N^r_{(2,3)}(n\cdot u)_{(2,3)} u_r^{(2,3)}  + (n\cdot u)_{(2,3)}^2 \right] ~,  
\end{aligned}
\end{equation}
\end{widetext}
and $N^r_{(2)}$ is given by \eqref{shift}, while $N^r_{(3)}=\sqrt{\varepsilon}$.
The first term is zero as computed by the interior metric. Now we can perform the integration over $\dd r$, and solve the constraint 
\begin{equation} \label{mass}
\begin{aligned}
M(R,t)=4 \pi \int_{R-\epsilon}^{R+\epsilon}r^2\sigma \delta(\chi_{(2,3)})\left[N^r_{(2,3)}(n\cdot u)_{(2,3)} u_r^{(2,3)} \right. \\ \left. + (n\cdot u)_{(2,3)}^2 \right]\dd r \,.
\end{aligned}
\end{equation}
where we simplified $1/\sqrt{1+\varepsilon}$, since is an overall constant factor.
First, one can easily show that $(n\cdot u)_{(2,3)}=-\Dot{T}_{(2,3)}$, while, using the first junction condition, $\Dot{T}_{(2,3)}$ can be entirely written with respect to $N^r_{(2,3)}$ and $\Dot{R}$.
We get
\begin{align} \label{first_israel}
&\dot{T}_{(2)}=\frac{N^r_{(2)}\dot{R}\pm_I\sqrt{(1+\varepsilon) \left[1+\varepsilon+\dot{R}^2-(N^r_{(2)})^2 \right]  }}{1+\varepsilon-(N^r_{(2)})^2} \,,  \\
&\dot{T}_{(3)}=\sqrt{\varepsilon}\dot{R}+\sqrt{(1+\varepsilon)(1+\dot{R}^2)}\,,
\end{align}
where in the interior the sign is forced to get $T_{(2)}$ with the same direction of $\tau$, while the sign choice $\pm_I$ for the exterior require a careful analysis, that we perform later.
Instead
\begin{align}
    & u_r^{(2,3)}=\frac{\Dot{R}}{1+\varepsilon}+\Dot{T}_{(2,3)} \frac{N^r_{(2,3)}}{1+\varepsilon}~.
\end{align}
So we get
\begin{align}
 & N^r_{(2,3)} (n\cdot u)_{(2,3)} u^{(2,3)}_r=-\frac{\Dot{T}_{(2,3)}N^r_{(2,3)} }{1+\varepsilon}\left( \Dot{R}+ N^r_{(2,3)}\Dot{T}_{(2,3)} \right)~,
 \\
 &(n\cdot u)^2_{(2,3)}=(\Dot{T}_{(2,3)})^2~.
\end{align}
What is left is to compute then the $4$-vector $\chi^\mu_{(2,3)}$, that allows to transform the Dirac delta as a function of $r$. Since $\chi^\mu$ is the spatial normal to $\Sigma_-$, it has to satisfy:

\begin{equation} \label{tan_normal}
    \begin{cases}
        & \chi^\mu_{(2,3)} \chi^{(2,3)}_\mu=1~,\\
        & \chi^\mu_{(2,3)} u_\mu^{(2,3)}=0 ~,\\
        &\chi^\mu_{(2,3)}=(A_{(2,3)}(r,t), B_{(2,3)}(r,t),0,0)~.
    \end{cases}
\end{equation}
We need to solve this system to find  $A_{(2,3)}(r,t)$, $B_{(2,3)}(r,t)$. For $\mathcal{M}_3$
\begin{align}
&   A_{(3)}(t,r) ={\dot{R}\sqrt{1 + \varepsilon} +\sqrt{\varepsilon} \sqrt{1+ \dot{R}^2}}  ~, \\
& B_{(3)}(t,r)= \sqrt{1 + \dot{R}^2} ~,
\end{align}
while for $\mathcal{M}_2$
\begin{align}
&   A_{(2)}(t,r) = \frac{\dot{R}\sqrt{1 + \varepsilon} \pm_I N_{(2)}^r \sqrt{1 - (N^r_{(2)})^2 + \dot{R}^2 + \varepsilon}}{1 - (N^r_{(2)})^2 + \varepsilon}  ~,  \\
& B_{(2)}(t,r)=\pm_I \sqrt{1 - (N_{(2)}^r)^2 + \dot{R}^2 + \varepsilon} \, .
\end{align}
In order to express the Dirac delta in terms of $r$, we need the transformation law between the two charts $\{t, r, \theta, \phi\}$ and $\{\tau, \chi, \theta, \phi\}$. It is given by
\begin{equation}
    \begin{cases}
        & \dd t = \frac{\partial t }{\partial \tau}\big|_{\chi} \dd \tau + \frac{\partial t}{\partial \chi}\big|_\tau \dd \chi =u^t \dd \tau + \chi^t \dd \chi  ~,  \\
        & \dd r = \frac{\partial r }{\partial \tau}\big|_{\chi} \dd \tau + \frac{\partial r}{\partial \chi}\big|_\tau \dd \chi =u^r \dd \tau + \chi^r \dd \chi~,
        \end{cases} \label{system}
\end{equation}
where the ${(2,3)}$ label is omitted for brevity, but this transformation is in general different between the interior and exterior. The Dirac delta transforms in general as
\begin{equation}
  \delta(\chi)=\frac{\dd r}{\dd \chi} \delta(r-R) ~. 
\end{equation}
The integration of the scalar constraint (as well as the diffeomorphism constraint) is carried out along the $t=const.$ hypersurface, which is in general discontinuous across the shell. Therefore, from \eqref{system} we get
\begin{equation}
 \frac{\dd r}{\dd \chi_{(2,3)}}=-u^r_{(2,3)} \frac{\chi^t_{(2,3)}}{u^t_{(2,3)}}+\chi^r_{(2,3)}=\frac{\sqrt{1 + \varepsilon}}{\Dot{T}_{(2,3)}}~.
\end{equation}
By plugging all these results in \eqref{mass} we get
\begin{equation}
\begin{aligned}
 &M(R,T_{(2)})= \\   &4 \pi \int_{R-\epsilon}^{R+\epsilon}r^2\sigma \delta(r-R) \frac{  \dot{T}(1 + \varepsilon) - N^r ( N^r \dot{T}+ \dot{R} ) }{\sqrt{1 + \varepsilon}}  \dd r \,.
 \end{aligned}
\end{equation}
Upon integrating along the radial direction, the integrand features functions that are discontinuous across $r = R$ and multiplied by a Dirac delta distribution. Consequently, adopting the standard convention for integrals involving discontinuities, we evaluate the term as
\begin{equation}
    \int F(r) \, \delta(r - R) \, \mathrm{d}r = \frac{F_+(R) + F_-(R)}{2}~.
\end{equation}
Carrying out the integration and taking the limit $\epsilon \to 0$,
\begin{equation} \label{mass_solution} 
\begin{aligned} 
M(R,T_{(2)})= &\frac{m(R,T_{(2)})}{2} \left[  \sqrt{1 + \dot{R}^2
} \right. \\  &\left. \pm_I \sqrt{1 { - R^2 \tilde{f}\left( \frac{2GM(R,T_{(2)})}{R^3}   \right) }
+\dot{R}^2 }\right] \,,
\end{aligned}
\end{equation}
where we have defined $m := 4\pi R^2 \sigma$ as the shell's rest mass. This relation corresponds to the second Darmois-Israel junction condition and acts as the evolution equation for the thin shell. Remarkably, this expression matches the one obtained for $\varepsilon = 0$ \cite{Bobula:2026zlq}. This surprisingly simple equation stems directly from the validity of Birkhoff's theorem in the effective framework, representing a unique feature of the class of effective models considered in this work \cite{Giesel:2025kdl}: the shell dynamics is determined entirely by the interior and exterior spacetimes. Specifically, if Birkhoff's theorem holds, the external spacetime at any given time $t$ must be described by an effective Schwarzschild metric with mass parameter $M(R,t)$. Consequently, the equations of motion for the thin shell necessarily coincide with those of an isolated thin shell whose exterior is determined by $M(R,t)$.

\subsection{Covariant energy-momentum conservation across the shell}
 Since an incoming energy flux is present from $\mathcal{M}_2$, the shell's proper mass $m = 4\pi \sigma R^2$ is not expected to be conserved. Specifically, if the incoming radiation flux carries negative energy, the inertial mass $m$ of the shell decreases as the gravitational mass $M(R,t)$ decreases. To verify this behavior, one must solve the energy-momentum conservation laws across the shell. The equation that must be coupled with the first Darmois-Israel junction condition and the previous relation is given by
\begin{equation}
D^b S_{ab}=-F_a^{(2)}+F_a^{(3)} \,,   
\label{bilancio}
\end{equation}
where $a, b = \{\tau, \theta, \phi\}$ denote the coordinates on $\Sigma_-$, and $S_{ab}$ is the energy-momentum tensor projected onto the thin shell. Here, $D^b$ represents the intrinsic covariant derivative on the shell, and $F_a^{(2,3)}$ is the net energy-momentum flux across the shell originating from $\mathcal{M}_{2,3}$. Clearly, since the interior spacetime is Minkowski, we have $F_a^{(3)} = 0$. We recall that
\begin{equation}
 S_{ab}=e^{\mu}_a e^\nu_b S_{\mu \nu}  \, ,
\end{equation}
where $S_{\mu\nu}$ is defined by $T_{\mu\nu} = S_{\mu\nu} \delta(\chi)$, and $e^\mu_a$ denotes the components of the embedding vectors on the surface $\Sigma$, with components given by $e^\mu_\tau = u^\mu$, $e^\mu_\theta = \delta^\mu_\theta$, and $e^\mu_\phi = \delta^\mu_\phi$. Consequently, a straightforward calculation—which yields identical results regardless of whether the exterior metric of $\mathcal{M}_2$ or the interior metric of $\mathcal{M}_3$ is employed—leads to
\begin{equation}
  S_{ab}=\mathrm{diag}(\sigma, 0, 0)  ~.
\end{equation}
It remains to write explicitly $F^{(2)}$, and make the covariant derivative of $S_{ab}$. We have
\begin{equation}
    F^{(2)}_a ={^{(2)}}e^\mu_a T_{\mu \nu}^{(2)}  \chi^\nu_{(2)} ={^{(2)}}e^\mu_a\rho_{(2)} n^{(2)}_\nu n^{(2)}_\mu  \chi^\nu_{(2)} ~,
\end{equation}
where $T_{\mu \nu}^{(2)}=\rho_{(2)} n_\mu^{(2)} n_\nu^{(2)}$. Let us examine this expression for a generic $\varepsilon>0$
\begin{align}
 & F^{(2)}_\tau =\frac{\rho_{(2)} \dot{T}_{(2)} (\dot{R} + N^r_2 \dot{T}_{(2)})}{\sqrt{1 + \varepsilon}} ~, \\
 & F^{(2)}_\theta=F^{(2)}_\phi=0~,
\end{align}
where $\rho_{(2)} = \frac{\partial_r M(r,t)}{4\pi r^2}$ represents the energy density of the dust. By explicitly computing the covariant derivative in \eqref{bilancio}, one obtains
\begin{equation}
    \frac{\dd (4 \pi R^2 \sigma) }{\dd \tau} = \dot{m}= \frac{\rho_{(2)} \dot{T}_{(2)} (\dot{R} + N^r_{(2)} \dot{T}_{(2)})}{\sqrt{1 + \varepsilon}} ~,
    \label{newcons}
\end{equation}
where $\rho_{(2)}=\frac{\partial_r M(r,T_{(2)})}{4 \pi r^2}\big|_{r=R}$ is the dust energy density on $R$.
This equation, combined with \eqref{sistema} (the first Darmois-Israel junction condition) and \eqref{mass_solution} (the second), completely determines the dynamics of the thin shell. At this stage, we can employ Lemos' trick. Indeed, the relations derived thus far correspond to the junction conditions for an exterior spacetime given by a generic dust collapse in the unbound case ($\varepsilon = \text{const.} > 0$). Since $\mathcal{M}_2$ must describe an ingoing flux of Hawking-like partners (i.e., an ingoing Vaidya spacetime with negative energy density), we must evaluate these equations in the limit $\varepsilon \to \infty$. 

Let us begin with the first Darmois-Israel junction condition. Employing Lemos' trick entails treating the effective Vaidya metric \eqref{invad} as the exterior spacetime and rewriting the first Darmois-Israel junction condition accordingly. The pullback of the exterior metric onto the thin shell hypersurface is described by the coordinate chart $\{V_{(2)}(\tau), R(\tau), \theta, \phi\}$, yielding the first Darmois-Israel junction condition
\begin{equation}
\begin{cases}
& R_2(\tau)=R_3(\tau)=R(\tau) \,, \\
& \theta_2=\theta_3=\theta\,, \\
&\phi_2=\phi_3=\phi \,, \\
&\left(1-{(N^r_2)^2} \right)\dot{V}_{(2)}^2-{2}\dot{V}_{(2)} \Dot{R}_2  =1 ~.
\end{cases}
\label{sistema1}
\end{equation}
It can be shown (see App. \ref{app_Vaiydyarho}), that the energy density $\rho_{(2)}$ after Lemos' trick reduces to
\begin{equation}
  \rho_{(2)}(V_{(2)},R)=\frac{\partial_v M(v)}{4 \pi R^2 \varepsilon}\bigg|_{v=V_{(2)}}~.  
  \label{newrho}
\end{equation}
Instead, $\Dot{T}_{(2)}$ and $\Dot{V}_{(2)}$ are related by
\begin{equation}
 \dot{V}_{(2)}=\sqrt{\varepsilon}\dot{T}_{(2)}-\frac{\Dot{R}}{\varepsilon} ~.
 \label{almostrans}
\end{equation}
Substituting \eqref{newrho} and \eqref{almostrans} in \eqref{newcons}, and sending $\varepsilon\rightarrow +\infty$, one gets
\begin{equation}
 \frac{\dd (4 \pi R^2 \sigma) }{\dd \tau} = \Dot{m}= \dot{V}_{(2)}^2 {\partial_{V_{(2)}} M(V_{(2)})}   ~.\label{finalcons}
\end{equation}
This yields the energy-momentum conservation equation for the thin shell with an exterior described by the effective collapsing Vaidya spacetime. Next, we turn to the second Darmois-Israel junction condition. This condition is the easiest to adapt via Lemos' trick because, as discussed previously, the corresponding expression is independent of $\varepsilon$ owing to Birkhoff's theorem. Consequently, the trick simply amounts to modifying the functional dependence of the mass parameters. As established in Sec. \ref{sec_nulldust}, the mass $M$ depends solely on $V_{(2)}$ in the limit $\varepsilon \to +\infty$; moreover, it follows from \eqref{finalcons} that the shell mass also satisfies $m = m(V_{(2)})$, as its evolution is entirely governed by $V_{(2)}$. This leads to
\begin{equation}
\begin{aligned}
   M(V_{(2)})=& \frac{m(V_{(2)})}{2} \left[  \sqrt{1 + \dot{R}} \right. \\  &\left. \pm_I \sqrt{1 {- R^2 \tilde{f}\left( \frac{2GM(V_{(2)}) }{R^3}  \right)} 
   +\dot{R}^2 }\right] \,.
\end{aligned}
\end{equation}
This closes the system for the thin shell evolution. We summarize it {as follows}
\begin{empheq}[left=\empheqlbrace\,]{align}
& \Dot{m} = \dot{V}_{(2)}^2 \partial_{V_{(2)}} M(V_{(2)}) \label{dotm} \,,  \\
& \left(1 - (N^r_2)^2 \right)\dot{V}_{(2)}^2 - 2\dot{V}_{(2)} \Dot{R}_2 = 1 \label{ciccia} \,, \\
& \begin{aligned}
  & M(V_{(2)}) = \frac{m(V_{(2)})}{2} \left[ \sqrt{1 + {\dot{R}^2 } } \right.  \\ 
  & \left. \pm_I \sqrt{1 -{ R^2 \tilde{f}\left( \frac{2GM(V_{(2)})}{R^3} \right)} + \dot{R}^2} \right] \,.
  \end{aligned} \label{cicciabuona}
\end{empheq}
Before concluding the subsection, we want to point out that if one solves \eqref{ciccia}, finds
\begin{equation} \label{Vcand}
\begin{aligned}
 \dot{V}_{(2)}=& \frac{1}{1 -{ R^2 \tilde{f}\left( \frac{2GM(V_{(2)})}{R^3} \right)} }\left( \dot{R} \right. \\  &\left.   \pm_I \sqrt{ 1 -{ R^2 \tilde{f}\left( \frac{2GM(V_{(2)})}{R^3} \right)} +\dot{R}^2 } \right) \,.   
\end{aligned}
\end{equation}
We already anticipated that the apparent ambiguity in the choice of the sign $\pm_I$ requires careful analysis, and this is the aim of the next subsection.
\subsection{Branch choice} \label{sec_sub_branchchoice}
 As for $\dot{T}_{(2)}$ in \eqref{first_israel}, a sign ambiguity arises in $\dot{V}_{(2)}$. The sign $\pm_I$ must be chosen such that $\dot{V}_{(2)} > 0$, ensuring that the shell motion is considered forward in time. 

We first focus on the collapse regime $\dot{R} < 0$. Here, we restrict our analysis to black hole solutions where the metric function $1 - R^2 \tilde{f}\left( \frac{2 G M(V_{(2)})}{R^3} \right)$ admits two real roots in terms of the areal radius $R$, corresponding to the inner ({trapping}) horizon $r_\mathrm{in}$ and the outer ({trapping}) horizon $r_\mathrm{out}$. This configuration occurs for masses {$M > M_\mathrm{ext} \sim  \, m_{\mathrm{Pl}}$} \cite{Kelly:2020uwj}. Specifically, since this function is proportional to the product of the null expansions, it is positive in the non-trapped regions and negative in the trapped region. Consequently, when the thin shell resides outside the outer horizon or within the inner horizon, we have
\begin{equation}
{ 1 - R^2 \tilde{f}\left( \frac{2 G M(V_{(2)})}{R^3}  \right) }>0 \,,
\end{equation}
and the condition $\dot{V}_{(2)} > 0$ requires selecting the $+$ sign. When the shell is within the trapped region, the term inside the square root can vanish, i.e.
\begin{equation}
    \Dot{R}^2= {-1 + R^2 \tilde{f}\left( \frac{2 G M(V_{(2)})}{R^3} \right)  \,,}
\end{equation}
and the radius $R$ at which this condition is satisfied is referred to as the \emph{branching radius} \cite{Gralla:2025gzl}. This may signal a branch change, which corresponds to adopting the $-$ sign in \eqref{Vcand}. Such a choice remains consistent with the requirement $\dot{V}_{(2)} > 0$, since both the numerator and the denominator in that expression become negative. Because the sole requirement for fixing the sign is ensuring $\dot{V}_{(2)} > 0$---a condition that can be fulfilled by both choices within the trapped region---the ambiguity must be resolved through the thin-shell equation of motion \eqref{cicciabuona}. To this end, one can employ a squared version of this equation, namely
\begin{equation} \label{RdotsquareV}
\begin{aligned}
   \dot{R}^2=   -1 + \frac{M(V_{(2)})^2}{m^2} + \frac{1}{2}R^2\tilde{f}\left( 
   \frac{2 G M(V_{(2)})}{R^3} \right)  \\ + \frac{ r^4 \tilde{f}\left( 
   \frac{2 G M(V_{(2)})}{R^3}  \right)^2 m(V_{(2)})^2 }{16 M(V_{(2)})^2}  ~, 
\end{aligned}
\end{equation}
to solve for $R(\tau)$, while employing \eqref{cicciabuona} as a control equation. If this condition is violated during the dynamics under a given sign choice, one is forced to change the sign dynamically. Within the trapped region, two distinct directions lead to a decreasing areal radius $R$: an inward direction and an outward direction---a characteristic feature of trapped geometries. A change of branch corresponds to a transition between these directions, causing the thin shell to shift from ingoing to outgoing while maintaining a decreasing $R$. It is worth noting that such a branch change does not typically occur for geodesic motion. The direction of the thin shell motion can be related to the choice of the sign $\pm_I$ by examining $\dot{V}_{(2)}$. In the $\pm_I = -$ case, as the inner horizon is approached such that $1 - R^2 \tilde{f}\left( \frac{2 G M(V_{(2)})}{R^3} \right) \to 0$, one finds $\dot{V}_{(2)} \to +\infty$ (with $\dot{R} < 0$ during the collapse phase). This implies that the thin shell moves toward the inner horizon in an outgoing manner, thereby approaching the boundary of the trapped region which marks the limit of validity of the $V_{(2)}$ chart. Conversely, for the $\pm_I = +$ case, $\dot{V}_{(2)}$ remains finite upon crossing $r_{\mathrm{in}}$, corresponding to an ingoing motion of the thin shell.

\section{Complete spacetime picture} \label{sec_completespacetime}
 Having established the necessary framework in the preceding sections, we are now positioned to synthesize these developments and present a detailed picture of black hole formation due to the thin shell collapse and its subsequent evaporation. Specifically, we will solve for the exact thin-shell dynamics and trapping-horizon trajectories, followed by a detailed analysis of the causal structure of the resulting spacetime. {Our results for the two evaporation law approaches introduced in Sec. \ref{sec_evaporationlaws}, which yield different qualitative pictures, are summarized in Figs. \ref{fig_RV} and \ref{fig_conf_both}. To obtain the numerical results, we have used libraries of \texttt{JULIA} programming language. }
\subsection{Solving for the evaporating thin-shell trajectory} \label{subsec_solvingfortrajectory}
 Using the relations $\frac{\dd R}{\dd V_{(2)}} = \frac{\dot{R}}{\dot{V}_{(2)}}$ and $\frac{\dd m}{\dd V_{(2)}} = \frac{\dot{m}}{\dot{V}_{(2)}}$, and defining $\mathcal{S} := \pm \sqrt{\dot{R}^2}$—where $\dot{R}^2$ is given by the right-hand side of Eq. \eqref{RdotsquareV}—we can recast the system in Eqs. \eqref{dotm}--\eqref{cicciabuona} into a $V_{(2)}$-dependent form. Furthermore, specializing to the black hole model described in Sec. \ref{sec_lqgdustcollapse}, we implement the substitution $1 - R^2 \tilde{f}\left( \frac{2GM(V_{(2)})}{R^3}\right) \to f_\mathrm{qS}(R) \big\vert{}_{M=M(V_{(2)})} =: f_\mathrm{qS}(V_{(2)},R)$. The resulting equations for numerical integration are given by
\begin{equation}  \label{sysnum}
\left\{
\begin{aligned}
&\frac{\dd R}{\dd V_{(2)}}=\frac{f_\mathrm{qS}(V_{(2)},R) \mathcal{S} }{ \mathcal{S} \pm_I \sqrt{f_\mathrm{qS}(V_{(2)},R)+\mathcal{S} ^2}} \,, \\
&\frac{\dd m}{\dd V_{(2)}} =\left(\frac{\mathcal{S} \pm_I \sqrt{f_\mathrm{qS}(V_{(2)},R)+\mathcal{S}^2}}{f_\mathrm{qS}(V_{(2)},R)}\right) \frac{\dd M(V_{(2)})}{\dd V_{(2)}} \,,
\end{aligned}
\right.
\end{equation}
where $M(V_{(2)})$ is obtained by converting the mass function $M(U_{(1)})$—determined by the evaporation laws introduced in Sec. \ref{sec_evaporationlaws} {and plotted in Fig. \ref{fig_MM_comparison}}—into the ingoing Vaidya-like mass function via the junction conditions presented in Sec. \ref{sec_hawking_israel}. By assuming a constant mass, the system \eqref{sysnum} can also describe thin-shell dynamics before the evaporation process begins. The sign of $\mathcal{S}$ is negative for a contracting thin shell and must be reversed to positive upon the shell's bounce at $\mathcal{S} = 0$. System \eqref{sysnum} remains valid as long as $m > 0$. As the solutions will demonstrate, the mass drops to zero soon after the quantum bounce of the shell, causing the thin shell to transition into a null shell with zero proper mass but non-zero energy. This behavior arises from the strong blue-shift effect experienced by the ingoing negative-energy radiation, which drastically reduces the rest mass $m$ of the shell in comparison to the more gradual decrease of the gravitational mass $M$. 

Because the shell becomes massless, the proper time $\tau${---the time parameter in the system in Eqs. \eqref{dotm}--\eqref{cicciabuona})---}ceases to be a well-defined parameter for describing its evolution, as the shell hypersurface is no longer timelike. This can be readily seen from \eqref{RdotsquareV}, where $\dot{R} \to +\infty$ as $m \to 0$. Although \eqref{RdotsquareV} and the second equation in \eqref{sysnum} break down in the null limit, one can still rely on the first equation of \eqref{sysnum} alongside the solution $M(V_{(2)})$ dictated by the evaporation law, given that the coordinate $V_{(2)}$ remains well-defined throughout the shell's evolution. Therefore, once the shell becomes null (which occurs inside the inner horizon), its null dynamics can be investigated by taking the limit $\mathcal{S} \to +\infty$ in the first relation of \eqref{sysnum}. This yields
\begin{equation} \label{shellnullv}
 \frac{\dd R}{\dd V_{(2)}}=\frac{f_\mathrm{qS}(V_{(2)},R)}{2}  \,,
\end{equation}
fully characterizing the null thin shell dynamics. Note that this corresponds to the radial velocity of an outgoing null ray in this effective Vaidya spacetime, indicating that, even at the effective level, a null thin shell retains a geodesic motion. Note that the right-hand side of \eqref{shellnullv} is positive as long as the shell remains in the non-trapped region. 

\subsection{Constructing conformal diagrams}{To illustrate the physically relevant properties of evaporating black hole spacetimes, it is useful to construct their Penrose-Carter (conformal) diagrams. Here, we provide the toolkit used to generate the diagrams in Fig.~\ref{fig_conf_both}, following the methodology introduced in \cite{Bobula:2024ywp, Bobula:2024chr}. These tools enable the numerical computation of conformal diagrams even for time-dependent, black-hole spacetimes. However, because the spacetimes under consideration comprise several distinct subregions, we adopt a hybrid approach: only the most physically relevant regions are computed numerically, while the remaining sections are plotted schematically on top of them.

We are primarily interested in numerically computing the trajectories of inner and outer horizons, as well as shell trajectories, on conformal diagrams. Within our framework, these regions are governed by the ingoing Vaidya-like line element of Eq.~\eqref{vaidya} with the metric function $f_\mathrm{qS}(v,r)$, namely
\begin{equation} \label{vaidya_conformal}
    \dd s^2 = -f_\mathrm{qS}(v,r)\,\dd v^2 + 2\,\dd v\,\dd r + r^2\,\dd \Omega^2.
\end{equation}
To construct the corresponding Penrose--Carter diagrams, we require double-null coordinates. Since $v$ is already a null coordinate, our task reduces to constructing an outgoing null coordinate $u$, where $u=\mathrm{const.}$ on null geodesics, valid across the region described by Eq.~\eqref{vaidya_conformal}. A suitable candidate for $\dd u$, obtained by imposing the radial null condition in \eqref{vaidya_conformal}, is 
\begin{equation} \label{du}
    \dd u := \frac{1}{2} h(v,r) f_\mathrm{qS}(v,r)\, \dd v - h(v,r)\, \dd r \,,
\end{equation}
where $h(v,r)$ is an initially arbitrary integrating factor introduced so that $\dd u$ satisfies the closedness condition $\dd(\dd u) = 0$, which is necessary for the existence of the coordinate $u$ \cite{Bobula:2024chr}. For \eqref{du}, this condition can be written as a first-order partial differential equation
\begin{equation} \label{diffu}
    \frac{1}{2} h \frac{\partial f_{\mathrm{qS}}}{\partial r} + \frac{1}{2} f_{\mathrm{qS}} \frac{\partial h}{\partial r} + \frac{\partial h}{\partial v} = 0 \,.
\end{equation}
We solve \eqref{diffu} for $h(v,r)$ using the method of characteristic curves. Parameterizing the solution surface by $(\mathdutchcal{s}, \mathdutchcal{m})$, we obtain the corresponding system of ordinary differential equations
\begin{equation} \label{ucons_system}
\left\{\begin{aligned}
    \frac{\dd v}{\dd \mathdutchcal{s}} &= 1 \,, \\
    \frac{\dd r}{\dd \mathdutchcal{s}} &= \frac{1}{2} f_\mathrm{qS} \,, \\
    \frac{\dd h}{\dd \mathdutchcal{s}} &= -\frac{1}{2} h \frac{\partial f_\mathrm{qS}}{\partial r} \,.
\end{aligned}\right.
\end{equation}
For convenience, we choose the initial conditions $v(0) = v_\mathrm{uin} = \mathrm{const.}$, $r(0) = \mathdutchcal{m}$, and $h(0) = h_\mathrm{uin}(\mathdutchcal{m})$ at $\mathdutchcal{s}=0$. Consequently, the differentials along the initial hypersurface $\mathdutchcal{s}=0$ satisfy $\dd r \big|_{\mathdutchcal{s}=0} = \frac{1}{2} f_\mathrm{qS} \, \dd \mathdutchcal{s} + \dd \mathdutchcal{m}$ and $\dd v \big|_{\mathdutchcal{s}=0} = \dd \mathdutchcal{s}$. Substituting these into \eqref{du} and integrating along $\mathdutchcal{s}=0$ yields
\begin{equation} \label{us0}
    u(0,\mathdutchcal{m}) = u_\mathrm{uin} - \int_{\mathdutchcal{m}_\mathrm{uin}}^{\mathdutchcal{m}} h(0,\tilde{\mathdutchcal{m}}) \, \dd \tilde{\mathdutchcal{m}} \,.
\end{equation}
That is, $u$ in \eqref{us0} serves as the second null coordinate, though it is initially defined only on the hypersurface $\mathdutchcal{s}=0$. Nevertheless, this is sufficient for our purposes, as we now demonstrate. We observe that the characteristic curves of the system \eqref{ucons_system} follow outgoing radial null geodesics, which is verified by noting that
\begin{equation}
    \frac{\dd r / \dd \mathdutchcal{s}}{\dd v / \dd \mathdutchcal{s}} = \frac{\dd r}{\dd v} = \frac{1}{2} f_\mathrm{qS} \,.
\end{equation}
Combined with our choice that $r = \mathdutchcal{m}$ on $\mathdutchcal{s}=0$, the $u$-coordinate at any point $(v,r)$ off the $\mathdutchcal{s}=0$ hypersurface is obtained by tracing that point—propagated either to the past or future—back to $\mathdutchcal{s}=0$ along its outgoing null geodesic. In other words, we can write $u(v,r) = u\bigl(0, \mathdutchcal{m}(v,r)\bigr)$
where $\mathdutchcal{m}(v,r)$ denotes the radial value at which the outgoing null geodesic passing through $(v,r)$ intersects the $\mathdutchcal{s}=0$ surface. In the above construction, the choice of $v_\mathrm{uin}$ determines the location of the $\mathdutchcal{s}=0$ hypersurface (a $v = \mathrm{const.}$ slice), while the function $h_\mathrm{uin}(\mathdutchcal{m})$ governs the visual appearance of the resulting conformal diagram. 

With the double-null coordinate pair $(u,v)$ at hand, we proceed to compactify them via
\begin{equation}
    \tilde{v} = \frac{1}{\pi} \arctan(\kappa_v v) \,, \qquad \tilde{u} = \frac{1}{\pi} \arctan(\kappa_u u) \,,
\end{equation}
where $\kappa_v$ and $\kappa_u$ are positive constants that likewise control the visual appearance of the diagram. The above construction suffices to generate the relevant spacetime regions discussed in Secs.~\ref{sec_sub_stefancomplete} and~\ref{sec_sub_schwinger_picture}, whose conformal diagrams are presented in Fig.~\ref{fig_conf_both}.
}

\begin{figure}
    \centering
    \includegraphics[width=1\linewidth]{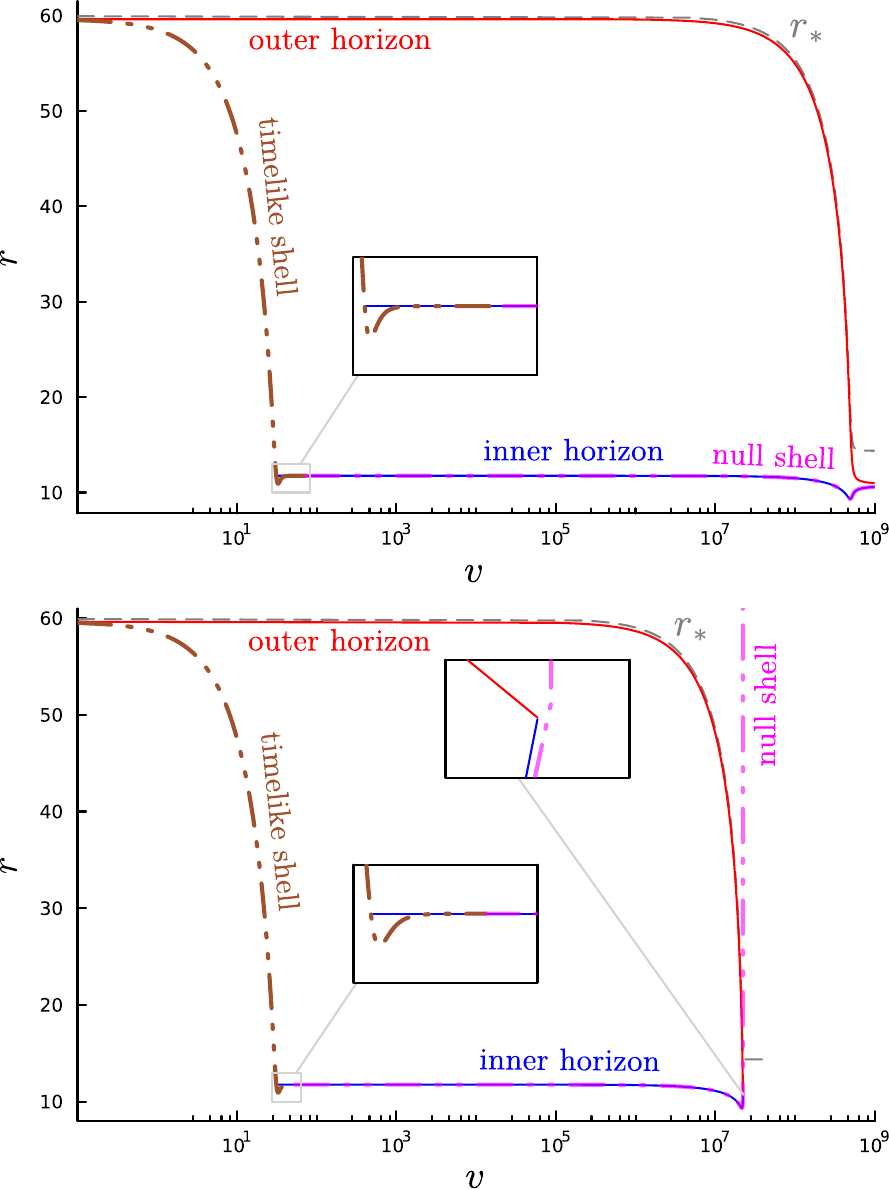}
    \caption{{ \emph{Top panel.}  Horizon and thin-shell trajectories in the $(v, r)$ plane obtained using the Stefan--Boltzmann evaporation law, plotting the areal radius $r$ against the null time coordinate $v$ (parameterized as $V_{(2)}$ in Eqs.~\eqref{sysnum} and \eqref{shellnullv}). After bouncing, the timelike thin shell (brown dash-dot-dot curve) rapidly transitions to a null shell, remaining trapped inside the inner horizon throughout late-stage evaporation. As $v \to \infty$, the outer (red) and inner (blue) horizons asymptotically merge. The $r_*$ surface (gray dashed curve) tracks closely along the outer horizon at all times. Parameters match those in Fig. \ref{fig_MM_comparison}, with $R(0) \approx r_\mathrm{out}(M(0))$ and $m(0) =  M(0)$. \emph{Bottom panel.} Analogous trajectories obtained under the Schwinger evaporation law, using the same curve styles and color coding as the top panel. Here, the outer and inner horizons merge at a finite time $v$. Once the horizons disappear, the null shell rapidly expands and crosses the $r_*$ surface, marking the end of the evaporation process. All model parameters are identical to those in the top panel.  }
    }
    \label{fig_RV}
\end{figure}

\subsection{Stefan-Boltzmann picture} \label{sec_sub_stefancomplete}
{Using the numerical setup established in Sec.~\ref{subsec_solvingfortrajectory}, we present the results of integrations of \eqref{sysnum} and \eqref{shellnullv} in the top panel of Fig.~\ref{fig_RV}. Here, we assume evaporation begins at $v = V_{(2)} = 0$—set at the crossing point of the timelike thin shell and the outer horizon (i.e., when the outer horizon forms)—with the same initial mass used in Fig.~\ref{fig_MM_comparison}. The initial mass $m$ of the thin shell is set equal to the mass $M$, yielding an initial marginally bound configuration. 

As shown in Fig.~\ref{fig_RV}, the $v$-timescale of the timelike thin shell prior to its transition to a null shell is much shorter than that of the null-shell phase, during which most of the gravitational mass evaporates. After the timelike-to-null transition occurs, the null shell resides in very close vicinity to the inner horizon. The inner and outer horizons appear to merge at the same areal radius—corresponding to the extremal black hole horizon radius—in the asymptotic limit $v \to \infty$. 

We present the corresponding conformal diagram in the left panel of Fig.~\ref{fig_conf_both}. In this diagram, the patch described by the ingoing Vaidya-like null dust encompasses the entire trapped region (bounded by the inner and outer horizons and a segment of the timelike shell trajectory). By contrast, the outgoing Vaidya-like null dust occupies only a tiny portion of the diagram due to our choice of functions and parameters governing its visual appearance. Throughout the evolution, the outer horizon remains timelike, whereas the inner horizon admits both timelike and spacelike segments. Finally, the dash-dotted curve represents the timelike thin shell prior to the onset of evaporation, continuing into the dash-double-dotted curve where evaporation takes place. The region to the left of the timelike—and subsequently null—shell trajectory is described by the Minkowski metric and bounded by the center of spherical symmetry at areal radius $r=0$. The white region to the right of the outgoing null dust patch and the dash-dotted segment of the thin-shell trajectory is described by the static black hole geometry $f_\mathrm{qS}$. Because a congruence of radial null geodesics reaches a finite areal radius as $v \to \infty$, the spacetime can be extended beyond this hypersurface. Along these finite-$r$ segments in the $v \to \infty$ limit (which coincide with the Cauchy horizon), we glue the spacetime to an extremal static black hole configuration. This extension introduces a new asymptotic region into which the null shell trajectory continues. 

The boundary of the causal past of future null infinity in the first asymptotic region defines the event horizon. Note that while the trajectory of the event horizon appears visually to coincide with the outer trapping horizon, this is not actually the case: the event horizon is by definition a null surface, whereas the outer trapping horizon remains timelike. Moreover, the location of the event horizon was only determined once the entire spacetime history is known, exemplifying the teleological nature of such horizons. Note that due to timelike character of the outer horizon, there exist null geodesics that escape trapped region and arrive at future null infinity in the first asymptotic region. Furthermore, the Cauchy horizon is plotted as an extension of future null infinity in the first asymptotic region.   

Based on the double-null coordinate construction in Sec.~\ref{sec_sub_conformaldiagrams}, the trajectories of the inner and outer horizons, as well as the timelike thin shell (including its constant-$u$ null shell extension), are numerically computed in the evaporating regime. Future and past null infinities are recovered in the limits $v \to \infty$ and $u \to -\infty$, respectively. All remaining features—including the event and Cauchy horizons, the $r=0$ surface, the pre-evaporation thin-shell trajectory, and the structures in the second asymptotic region—are drawn schematically.

}

\subsection{Schwinger picture} \label{sec_sub_schwinger_picture}
{Similarly as in Sec. \ref{sec_sub_stefancomplete} we solve the systems introduced in Sec. \ref{subsec_solvingfortrajectory} for $M(v)$ obtained for the Schwinger approach to evaporation law, with the same initial conditions. The results are presented in bottom panel of Fig. \ref{fig_RV}.

The initial $v$-time evolution of the thin shell—and thus the early stage of the overall picture—is qualitatively the same as in the previous case. However, crucial differences emerge in the final stages. The merger of the outer and inner horizons in finite $v$-time introduces a key qualitative change: the null shell rapidly propagates towards $r \to \infty$ after the disappearance of the horizons.
}

{In the bottom panel of Fig. \ref{fig_RV},} the expanding null thin shell approaches the $r_*$ surface in finite $v$-time following the disappearance of the trapping horizons. This $v$-time marks the termination of evaporation, with the remnant gravitational energy $M(v) \sim  \, m_{\mathrm{Pl}}$ being carried entirely by the thin shell. The evaporation terminates because, once the shell crosses $r_*$, the region beneath the shell is vacuum: since the shell is null, it lies on the boundary of the causal shadow of the crossing point. Consequently, the $r_*$ surface lies in a vacuum beneath the null shell, preventing further evaporation. Given that the shell is null and outgoing---as implied by \eqref{shellnullv} since the shell propagates in a non-trapped region---this result is independent of the specific choice of the timelike evaporation surface beyond $M_{\mathrm{ext}}$ considered in Sec.\ \ref{sec_hawking_israel}, as the shell will in any case reach the $r_*$ surface in finite $v$-time due to its outgoing motion. {The discussion so far corresponds to the case where $r_*$ remains very close to the outer horizon for as long as that horizon persists. A natural question is whether this effect—namely, the emission of a remnant null shell—would still occur if $r_*$ were located far from the outer horizon during the relevant stages of the evolution, as in the seminal work of Hayward \cite{Hayward:2005gi}, where the analogous surface $r_*$ was held constant throughout the entire evaporation process. We expect an affirmative answer to this question: as shown in Fig.~\ref{fig_RV}, the null shell expands rapidly to infinity after the horizons disappear, whereas evaporation is a slow process with respect to the $v$ or $u$ coordinates. Note also that in the exterior of the null shell, the rate of particle production drops rapidly at large $r$, as seen in Fig.~\ref{fig_dNdudr}. Therefore, any particle production in the exterior of the horizonless null shell—that is, the null shell propagating in the final horizonless spacetime—is negligible.} To conclude, we point out that even though locating the evaporating surface far from the horizon could be mathematically interesting, it lacks physical motivation: indeed, even though the formalism allows to place such boundary everywhere, since the Vaidya solution describing Hawking-like partners violates the null energy condition (just like in the Hawking computations), it is more reasonable that such solution lies almost entirely within the trapped region. 

{The right panel of Fig.~\ref{fig_conf_both} shows the resulting conformal diagram, which differs crucially from that obtained under the Stefan–Boltzmann evaporation law. Because the trapping horizons merge in finite $v$-time, all radial null geodesics traversing the trapped region eventually expand to infinite areal radius and reach future null infinity. Consequently, unlike in the previous case, no spacetime extension is possible beyond $v \to \infty$. As a result, neither an event horizon nor a Cauchy horizon forms, and the entire spacetime remains globally hyperbolic. Aside from this difference, the remaining building blocks of the diagram—such as the locations of the Minkowski and static black hole regions—are analogous to those in the previous case discussed in Sec. \ref{sec_sub_stefancomplete}. }

{As in Sec.~\ref{sec_sub_stefancomplete} for the previous diagram, the numerically computed objects are the trajectories of the trapping horizons, the null infinities, and the timelike and null shell trajectories, while the remaining objects are drawn schematically. }

{We also comment on the branching radius discussed in Sec.~\ref{sec_sub_branchchoice}. The mass considered in Figs.~\ref{fig_RV} and \ref{fig_conf_both} is too small for branching to occur, both in the evaporating case and in the static black hole formation limit where evaporation is neglected. For larger masses where branching does occur, the qualitative features of the resulting causal structure appear to be unchanged, however, extracting conformal diagrams becomes numerically more demanding due to the longer evaporation timescales. Furthermore, in the Schwinger case, global hyperbolicity is guaranteed by the mass evaporating below the extremal limit within a finite $u$- or $v$-time. Consequently, this global causal structure should also persist for larger masses. }

\label{sec_sub_conformaldiagrams}
\begin{figure*}
    \centering
    \includegraphics[width=1\linewidth]{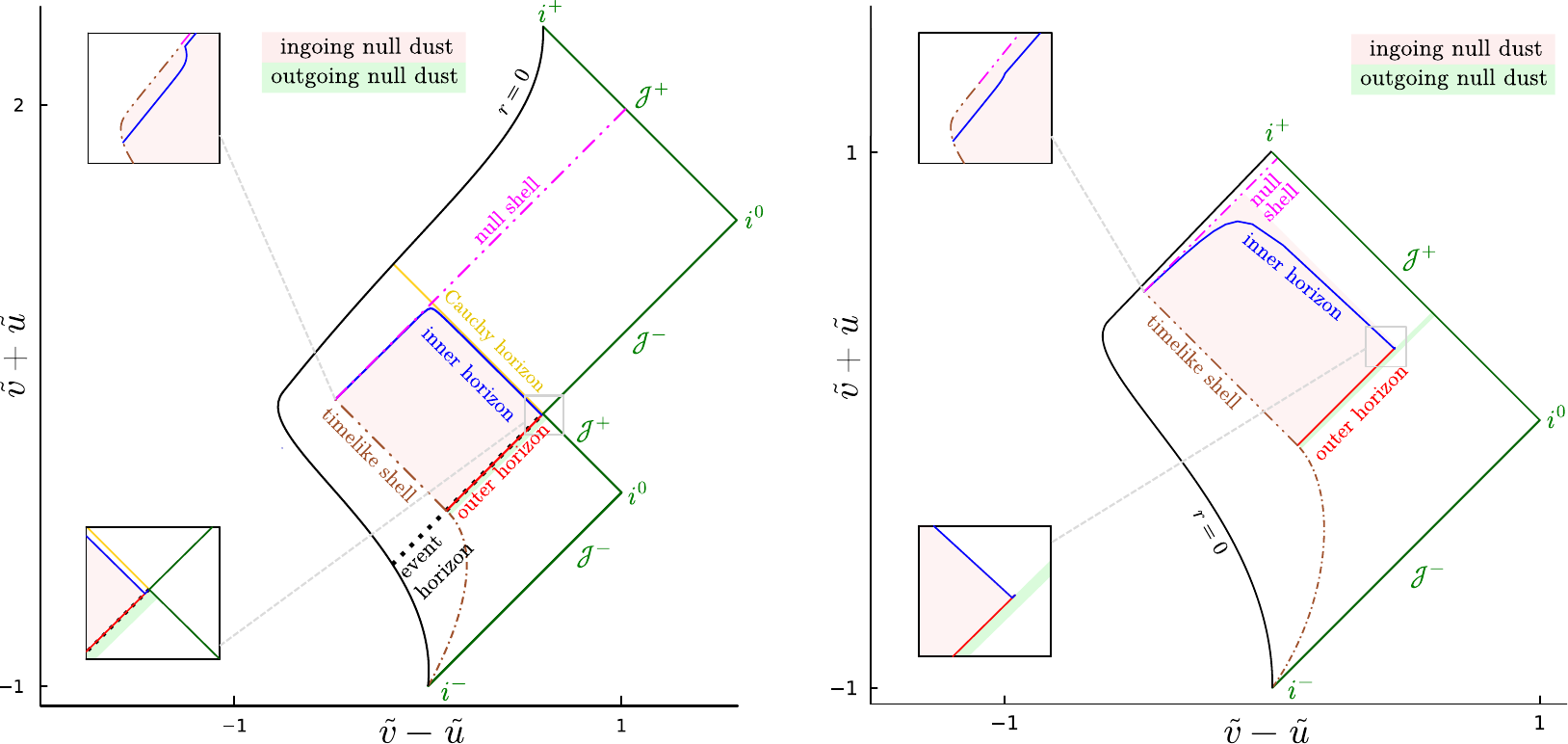}
    \caption{{\emph{Left panel.} Conformal diagram of black hole formation from thin-shell collapse and subsequent evaporation via the Stefan–Boltzmann law. Ingoing and outgoing null dust are emitted from the $r_*$ surface (not shown), located in the immediate vicinity of the outer horizon. The inner and outer trapping horizons appear to merge asymptotically at infinite time as measured by observers in the lower asymptotic region. Following its bounce, the timelike thin shell rapidly transitions into a null shell. Once evaporation concludes, the spacetime is extended by joining a static extremal black hole region, through which the null thin shell propagates to future null infinity. Both Cauchy and event horizons are present in this geometry. This plot uses the same parameters as Fig. \ref{fig_RV}, with $v_\mathrm{uin} = \mathdutchcal{m}_\mathrm{uin} \approx 32.83$ (this $v$-time marks timelike thin shell bounce), $u_\mathrm{uin}=1.5$, {$\kappa_v = 10^{-7}$, $\kappa_u = 1$} and $h_\mathrm{uin}(\mathdutchcal{m})= 3 \, \mathrm{sech}[0.6(\mathdutchcal{m}-30)]$. \emph{Right panel. } Analogous conformal diagram for black hole evaporation via the Schwinger approach. In contrast to the diagram on the left, the inner and outer horizons merge and disappear in finite time as measured by asymptotic observers. The remnant null shell propagates to future null infinity. The entire process takes place within a single asymptotic region, yielding a globally hyperbolic spacetime free of event and Cauchy horizons. This plot uses the same parameters as Fig. \ref{fig_RV}, with $v_\mathrm{uin} = \mathdutchcal{m}_\mathrm{uin}  \approx 32.83$ (this $v$-time marks timelike thin shell bounce) {and the same $u_\mathrm{uin}$, $\kappa_v$, $\kappa_u$, $h_\mathrm{uin}(\mathdutchcal{m})$} as for the left panel.  } }
    \label{fig_conf_both}
\end{figure*}

\subsection{Tangential pressure on the evaporation surface}
 To describe all the features of the overall spacetime, we now return to the evaporation surface $r_*$. As anticipated in Sec.\ \ref{sec_hawking_israel}, ensuring a continuous mass function at $r_*$ requires the absence of any surface density on the dynamical boundary. This can be verified by applying the second Darmois-Israel junction condition in the Hamiltonian formulation to this boundary, following the same steps introduced in this section for the computation on the $\Sigma_-$ boundary. For subsequent purposes, we assume that the energy-momentum tensor on the shell contains not only a surface density $\sigma$ but also a tangential pressure $p_\Sigma$. A direct computation yields
\begin{equation}
\begin{aligned}
\delta M= 4 \pi \int_{r_*-\epsilon}^{r_*+\epsilon}r^2\sigma\delta(\chi_{(1,2)})\left[N^r_{(1,2)}(n\cdot u)_{(1,2)} u_r^{(1,2)} \right. \\ \left. + (n\cdot u)_{(1,2)}^2 \right]\dd r \,,
\end{aligned}
\end{equation}
where $\delta M$ denotes the mass jump across the boundary. One immediately recognizes that $\sigma = 0$ implies $\delta M = 0$. Furthermore, applying Lemos' trick to the trivial identity $\delta M \equiv M(r_*, T_{(1)}) - M(r_*, T_{(2)}) = 0$ yields $M(U_{(1)}) = M(V_{(2)})$. As demonstrated in App. \ref{appendixpress}, the tangential pressure does not enter the calculation of the second Darmois-Israel junction condition.

As noted also in \cite{Hayward:2005gi} for a toy model of nonsingular black hole formation and evaporation, the absence of surface energy density does not imply necessary absence of tangential pressure on such a boundary. Instead, it is shown in App. \ref{appendixpress} that tangential pressure $p_\Sigma$ is needed, to get a consistent energy-momentum conservation across the boundary. Specifically, a short computation gives
\begin{equation}
   p_{\Sigma}=\frac{1}{8 \pi G r_*} \frac{\frac{\dd U_{(1)}}{\dd V_{(2)}}\dot{U}_{(1)}^2-\dot{V}_{(2)}^2}{\frac{\dd U_{(1)}}{\dd V_{(2)}} \dot{U}_{(1)}^2}~, 
\end{equation}
where $\frac{\dd U_{(1)}}{\dd V_{(2)}}$, given by , depends on the specific evaporation law considered. The plot of $p_\Sigma$ for the two cases is shown in Fig. \ref{fig:placeholder}.

{We also note that for a larger initial mass at the onset of evaporation, a negative tangential pressure could develop at $r_*$, for the Schwinger approach to evaporation. However, we interpret this feature as an artifact of our specific choice of $r_*$, that is, if one instead chooses a constant-radius surface throughout the entire evaporation process, as in Hayward's work \cite{Hayward:2005gi}, the tangential pressure vanishes at all times (see App. \ref{appendixpress}). }

\begin{figure}
    \centering
    \includegraphics[width=1\linewidth]{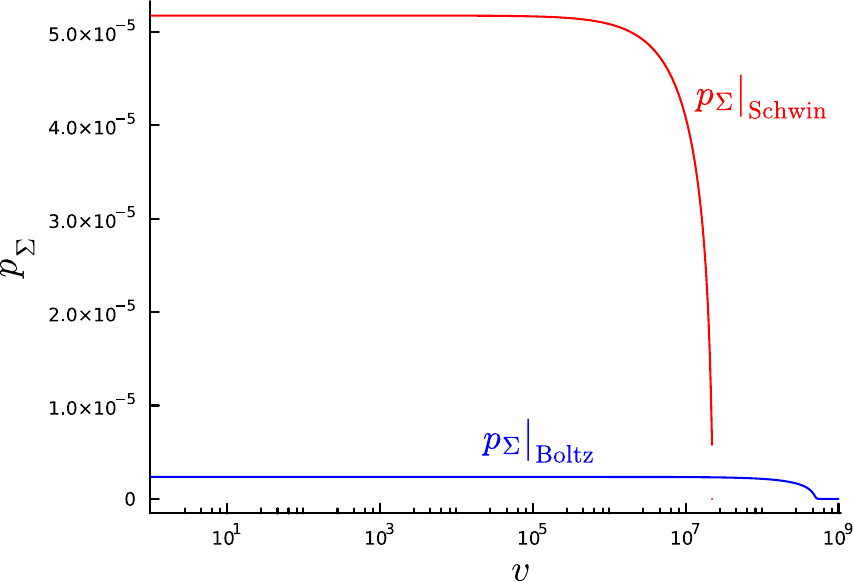}
    \caption{{Tangential pressures $p_\Sigma(v)$ for black hole evaporation corresponding to Figs.~\ref{fig_RV} and \ref{fig_conf_both}, evaluated on the junction surface $r_*$ from which outgoing and ingoing null dust emerge. For evaporation governed by the Stefan–Boltzmann law, the pressure asymptotically approaches zero as most of the mass evaporates. For evaporation via the Schwinger mechanism, the pressure similarly decreases before jumping to zero once $r_*$ becomes a constant-radius surface following the disappearance of the trapping horizon, leaving a brief zero-pressure interval until evaporation terminates in $v$-time.   }
    }    
    \label{fig:placeholder}
\end{figure}
\section{Comparison with other models} \label{sec_comparison}

The model presented in this work provides a comprehensive picture of stellar collapse—modeled as a thin dust shell—black hole formation, and effective-level evaporation. {Here, we compare the model to earlier literature.}

Most effective models, rooted either in LQG or several alternatives, developed so far describe either only stellar collapse---both homogeneous (see e.g. \cite{Li:2025kou,Ou:2025bbv,Schmitz:2020vdr,Kelly:2020lec,Han:2023wxg,Giesel:2023hys,Cafaro:2024vrw,Alonso-Bardaji:2023qgu}), and inhomogeneous (e.g. \cite{Husain:2022gwp,Cipriani:2024nhx,Fazzini:2026uqp,Fazzini:2025hsf,Giesel:2026pjj,Liu:2025fil})---or only the vacuum solution (e.g. \cite{Gambini:2008dy,Gambini:2013ooa, Belfaqih:2024vfk,Alonso-Bardaji:2022ear,Zhang:2025ccx}) or its evaporating counterpart (e.g. \cite{Hayward:2005gi,Frolov:2017rjz,Belfaqih:2025eak,Belfaqih:2026qgj,Alonso-Bardaji:2025udt,Alonso-Bardaji:2025qft,Bonanno:2006eu}). In contrast, {our strategy and} the junction conditions implemented here at the effective level yield a unified, complete picture of {the entire process of dynamical formation and evaporation}, resulting in a globally hyperbolic spacetime\footnote{This was achieved for the Schwinger approach to the evaporation process.}. However, prior models aiming to provide a complete picture do exist \cite{Hayward:2005gi, Hossenfelder:2009fc, Boyanov:2025pes, Gralla:2025gzl, DiFilippo:2025kzh}. Some of these \cite{Hayward:2005gi, Hossenfelder:2009fc} either lack dynamical control over black hole formation or rely on Einstein's field equations with exotic sources in certain regions. A second category of models \cite{Boyanov:2025pes, Gralla:2025gzl, DiFilippo:2025kzh} do not account for both ingoing and outgoing fluxes—typically restricting their analysis to the ingoing sector—and additionally lacks a well-motivated evaporation law capable of sustaining evaporation beyond the extremal horizon configuration. Furthermore, unlike many previous constructions argued to provide a consistent picture while {neglecting the {Hawking-like radiation}, our construction avoids the classical white hole instability---typical of models with macroscopic anti-trapped regions \cite{Eardley:1974zz}.
The dynamical model considered here is consistent with Ashtekar-Paw{\l}owski-Singh loop quantum cosmological dynamics \cite{Ashtekar:2006rx, Ashtekar:2006uz, Ashtekar:2006wn}. When {Hawking-like radiation} is neglected and a homogeneous, isotropic dust profile is assumed, the dynamics yield a bouncing cosmology that resolves the cosmic singularity. Notably, this feature is absent (or the corresponding cosmological dynamics is not presented) in alternative proposals—specifically static nonsingular black hole models with no Cauchy horizons (see e.g. \cite{Belfaqih:2024vfk,Alonso-Bardaji:2022ear, Zhang:2024ney, Calza:2025mrt}).

\section{Summary and conclusions} 
\label{sec_discussion}
The model presented in this paper consists in an application of the Hamiltonian formulation of the Israel junction conditions at the effective level, originally introduced in \cite{Fazzini:2025nse}. Specifically, stellar collapse, black hole formation, and evaporation are modeled using a timelike dust thin shell for the collapsing body, ingoing null dust with negative energy to represent the flux of Hawking-like partners (yielding an ingoing effective Vaidya metric), and outgoing null dust with positive energy for the {Hawking-like radiation} flux (following Hiscock's approach \cite{Hiscock:1980ze, Hiscock:1981xb}). The only external input to the model is the evaporation law, for which we consider two distinct prescriptions: the standard evaporation law based on the Stefan-Boltzmann approach to {Hawking-like radiation} \cite{Page:1976df}, and one rooted in the gravitational analogue of the Schwinger effect \cite{Wondrak:2023zdi}. {The choice of a timelike thin shell to model black hole formation is motivated not only by its mathematical tractability but also by the results of \cite{Hellaby:1994bj,Fazzini:2023ova,Fazzini:2026uqp}. In particular, these studies showed that, for Gaussian-like initial density profiles (or more generally initial density profiles with an upward-sloping portion) the collapsing star develops shell-crossing singularities, and a natural extension beyond them is provided by the Darmois-Israel dust thin shells \cite{Fazzini:2025nse, Bobula:2026zlq}.}

The dynamical spacetime can be divided into three main regions: the interior of the collapsing thin shell, which is given by classical Minkowski spacetime; an intermediate region described by the ingoing Vaidya metric; and an external region described by the outgoing Vaidya metric. The timelike thin shell separates the first two regions, whereas an evaporating surface with zero surface density and non-vanishing tangential pressure---situated slightly outside the dynamical trapping horizon---separates the intermediate and external regions. The entire spacetime geometry is derived by systematically imposing two sets of Darmois-Israel junction conditions within the Hamiltonian framework.

The Vaidya spacetimes are derived from an underlying LQG-inspired effective model \cite{Giesel:2026pjj} via Lemos' trick, which allows one to recover null dust from the equations of motion and the metric of timelike dust, analogously to the classical theory \cite{Lemos1992,Hellaby:1994bj}.

The resulting spacetime picture depends heavily on the chosen evaporation law. In the Stefan-Boltzmann case, the trapped region bounded by the inner and outer horizons progressively shrinks due to evaporation, tending asymptotically to zero. The thin shell remains confined within the horizons during its evolution---consisting in a collapse, bounce and expanding phase---eventually becoming a null shell that reaches future null infinity $\mathcal{I}^+$ of a new asymptotic region with a Planckian gravitational mass. {Both event and Cauchy horizons} persist in this picture. In contrast, under the Schwinger approach, the inner and outer horizons merge in finite proper and coordinate time and subsequently disappear {leaving no event and Cauchy horizons}. Here too, the shell becomes null following the bounce; once the horizons vanish, the shell---carrying a residual Planckian gravitational mass---propagates outward through a non-trapped region, ultimately reaching $\mathcal{J}^+$. The whole spacetime dynamics in this case consists of a single asymptotic region.

In the Schwinger case, the complete spacetime is entirely free from macroscopic or microscopic white hole regions, thereby automatically avoiding the well-known Eardley instability \cite{Eardley:1974zz}, which typically plagues models featuring macroscopic anti-trapped regions. Furthermore, the spacetime is free from physical singularities, from Cauchy horizons, and is globally hyperbolic. This dynamical geometry also provides a novel framework for investigating the mass-inflation instability and its fate in quantum-gravitational black hole models. Indeed, the absence of a Cauchy horizon alleviates the mass-inflation problem characteristic of stationary or dynamical spacetimes harboring inner horizons \cite{Ori:1991zz,Poisson:1990eh}---around which typically ingoing null perturbations are infinitely blue-shifted.
Nevertheless, a finite mass accumulation is not strictly excluded by the model, manifesting in the form described in \cite{Bonanno:2022jjp,Carballo-Rubio:2024dca}. Further analysis will clarify whether this phenomenon can indeed occur within the present framework and, if so, how to address it.

By incorporating {Hawking-like radiation}, the model successfully avoids the formation of inter-universal wormholes---a common feature of many LQG-inspired effective models---thereby in principle enabling the recovery of the full information associated with the collapsed body. However, the absence of an event horizon and the global hyperbolicity of the spacetime{---obtained within the Schwinger approach to evaporation law---}might not automatically resolve the standard information-loss paradox, which demands not only that information is shielded from physical singularities, but also that the total quantum system remains in a pure state \cite{Page:1993wv}. To attain a pure state and recover the information, the remnant's geometry must encode the information concerning the entire collapse history \cite{Bianchi:2026ytw}. This might not be fully achieved within the present effective framework, necessitating quantum-gravitational effects beyond the semiclassical treatment. The need for a complete quantum treatment near the Planckian regime for the gravitational mass ($M \sim m_{\mathrm{Pl}}$) is also motivated by another consideration: the model heavily relies on the validity of the effective theory throughout the entire spacetime dynamics. Nevertheless, loop quantum cosmology insights about the interior region \cite{Rovelli:2013zaa} indicate that a full quantum treatment becomes indispensable when the mass approaches the Planck scale. From this perspective, the results presented here should be regarded as an initial step in this direction. Consequently, although a black-to-white hole transition like the one presented in \cite{Han:2023wxg} is excluded at the effective level, a quantum tunneling of the geometry \cite{DAmbrosio:2020mut,Dona:2024rdq,Rovelli:2024sjl,Dona:2025snr} remains in principle possible once a rigorous quantum-gravitational treatment of the remnant is implemented. Additionally, the adiabatic approximation is applied to the evaporation laws, assuming their validity throughout the entire process. While this approach {might} fail as the black hole mass approaches the Planck mass—where the evaporation rate might change dramatically—accounting for non-adiabatic effects in the final stages is deferred to future work. 

Although the matter content of the spacetime is modeled primarily by a timelike dust thin shell, the framework is general enough to accommodate more sophisticated collapse scenarios, such as the Oppenheimer-Snyder dust collapse, Oppenheimer-Snyder collapse with a thin shell modeling the formation of shell-crossing singularities after the matter bounce \cite{Bobula:2026zlq} in the dust or pressure fluid case, or even more exotic configurations like collapse in a de Sitter environment.  Investigating these directions constitutes the goal of future work.

{The presented model exemplifies a scenario in which backreaction is a necessary ingredient for consistent black hole evolution. If backreaction were neglected, the global hyperbolicity of the picture would be lost—as is also the case for recent quantum gravity-inspired black hole formation models \cite{Husain:2021ojz, Lewandowski:2022zce, Bonanno:2023rzk, Bueno:2024eig}. This situation, which is valid across broad approaches to quantum gravitational black hole formation, suggests that backreaction can serve as a predictability-preservation mechanism. In this regard, the question arises whether the presented LQG-inspired framework is unique among alternative approaches in achieving such consistency. We expect that it is likely not, as the crucial ingredient is complete mass evaporation in finite $v$-time. }

Our framework opens new avenues for predicting observational signatures beyond classical general relativity. While typical black hole interiors remain concealed behind event horizons—making direct probing theoretically impossible for external observers—the globally hyperbolic spacetime (obtained within the Schwinger approach to evaporation) features no event horizons. In the Boltzmann approach instead only a portion of the trapped region is exposed to external observers, due to the timelike nature of the dynamical trapping horizon. Consequently, signals from the interior can, in principle, escape and reach external observers. In particular, our model predicts that a Planckian-mass null shell propagates to future null infinity. Beyond its dynamics, the model's kinematic properties are equally promising. Specifically, analyzing light propagation within this spacetime—following the approaches of \cite{Cardoso:2023guh, Arrechea:2026hhl}, which predicted highly energetic bursts for such spacetimes—would represent a vital step beyond the toy models employed in previous studies.

\section*{Acknowledgements}
F.F. thanks Eugenio Bianchi and Alfio Bonanno for useful discussions. M.B. acknowledges support from the Polish National Science Centre (Narodowe Centrum Nauki, NCN) under SONATINA 10 Grant No. 2026/60/C/ST2/00157.

\appendix

\section{Vaidya energy density}
\label{app_Vaiydyarho}
     In order to recover \eqref{newrho}, we start from the coordinate transformation LTB$\rightarrow$Vaidya:
\begin{equation}
 v=\frac{t}{\sqrt{\varepsilon}}+\frac{r(x,t)}{\varepsilon}  ~, \label{ltbvad}
\end{equation}
and considering $\varepsilon$ large (not yet taking the limit)
\begin{align}
&r(x,t)\rightarrow \sqrt{\varepsilon}(a(x)-t)~,\\
    &\frac{\partial r(x,t)}{\partial x}\rightarrow ({\partial_{x}} a(x))\sqrt{\varepsilon}~,\\
    & v\rightarrow \frac{a(x)}{\sqrt{\varepsilon}}~.
\end{align}
This means
\begin{equation}
  \frac{\partial v}{\partial x}\bigg|_t=\frac{\partial_ {x} a}{\sqrt{\varepsilon}}=\frac{\partial r}{\partial x}\bigg|_t\frac{1}{\varepsilon}  ~.\label{a6}
\end{equation}
Now, we can look at \eqref{ltbvad} as a coordinate transformation from $\{t, x \}$ to $\{v,t \}$ in the limit $\varepsilon\rightarrow+\infty$. Indeed, in such limit $v$ becomes a function only of $x$. If you one the jacobian of the transformation, it is diagonal, and find
\begin{equation}
\frac{\partial x}{\partial v}\bigg|_t=\frac{1}{\frac{\partial v}{\partial x}\big|_t} ~. 
\end{equation}
Then 
\begin{equation}
 \frac{\partial M(v,t)}{\partial v}\bigg|_t=\frac{\partial M}{\partial x} \frac{\partial x}{\partial v}\bigg|_t=\varepsilon \frac{\frac{\partial M}{\partial x}}{\frac{\partial v}{\partial x}\big|_t}~. \label{vaidyamass}
\end{equation}
We can use this transformation law to write $\rho$ from LTB to Vaidya: starting from
\begin{equation}
\rho_{(2)}=\frac{{\partial_{x}} M(x)}{4 \pi r^2 \partial_{x} r}
\end{equation}
and then changing coordinates to the $\{v, t\}$ ones, using \eqref{a6} and \eqref{vaidyamass}
\begin{equation}
  \rho_{(2)}(v,t)=\frac{\partial_v M(v)}{4 \pi r(t,v)^2 {\varepsilon}}~,
\end{equation} 
where we justified the dependence of $M$ only from $v$ in Sec. \ref{sec_nulldust}. Then, pass to the coordinates $\{v,r \}$
\begin{equation} \label{rhov}
   \rho_{(2)}(v,r)=\frac{\partial_v M(v)}{4 \pi r^2 {\varepsilon}}~.
\end{equation}
Note that this does not coincide with $T_{vv}$ for null dust, but with ${^{LTB}}T_{tt}$ expressed in $\{v,r, \theta, \phi\}$ coordinates. To get the Vaidya energy-momentum tensor in these coordinates, we can first pass to the energy-momentum tensor in PG-like coordinates---a simple computation gives ${^\mathrm{PG}}T_{tt}={^\mathrm{LTB}}T_{tt}$---and then to advanced Eddington-Finkelstein coordinates, using
\begin{equation}
 v=\frac{t}{\sqrt{\varepsilon}}+\frac{r}{\varepsilon}  ~, \label{ltbvad1}
\end{equation}
to get
\begin{equation} \label{Tvv}
T_{vv}=\frac{\partial t}{\partial v}\bigg|_r\frac{\partial t }{\partial v}\bigg|_r {^\mathrm{PG}}T_{tt}=\frac{\partial_v M(v)}{4 \pi r^2} \,.
\end{equation}
This is the classical null dust energy density---the only non-vanishing component of the Vaidya energy momentum tensor in $\{v,r,\theta,\phi\}$---, that, as in the massive dust case, retains its form even at the effective level.\footnote{Interestingly, a recent covariant effective model~\cite{Zhang:2026tjc}, which appeared while this work was in preparation, reached the same conclusion following a completely different approach.} 

\section{Tangential pressure on $r_*$}
\label{appendixpress}
 In this appendix, we establish two main results. First, we show that the tangential pressure does not contribute to the second Darmois-Israel junction condition on the boundary $r_*$, second, we demonstrate that it is instead necessary to satisfy covariant energy-momentum conservation across $r_*$ itself. 

We begin by considering the two spacetimes $\mathcal{M}_{(1,2)}$ prior to the application of Lemos' trick. {They read (see Sec. \ref{sec_nulldust})
\begin{align}
 \dd s_1^2=& -\frac{1}{1+\varepsilon_1}\left(1-r^2 \tilde{f}\left( \frac{2 G M(t_1, r)}{r^3} \right) \right)\dd t_1^2 \notag \\
 &+\frac{\dd r^2}{1+\varepsilon_1} +\frac{2}{1+\varepsilon_1}N^r\dd r \dd t_1 + r^2 \dd \Omega^2~, \\
 \dd s^2_2=& -\frac{1}{1+\varepsilon_2}\left(1-r^2 \tilde{f}\left( \frac{2 G M(t_2,r)}{r^3} \right) \right)\dd t_2^2  ~\notag \\
 &+\frac{\dd r^2}{1+\varepsilon_2}+\frac{2}{1+\varepsilon_2}N^r\dd r \dd t_2 + r^2 \dd \Omega^2 ~.
\end{align}}
{Let us now focus on} the contribution from the tangential pressure term. We start by assuming an energy-momentum tensor on $r_*$ given by
\begin{equation}
 T^{shell}_{\mu \nu}=(\sigma u_\mu u_\nu + p_{\Sigma}(g_{\mu \nu}-\chi_\mu \chi_\nu+u_\mu u_\nu ))\delta(\chi)  ~,
\end{equation}
where we set $\sigma=0$. The contribution of the pressure term to the Hamiltonian constraint
\begin{equation}
  \mathcal{H}^\mathrm{shell}=\frac{4 \pi r_*^2}{\sqrt{1+\varepsilon}}n^\mu n^\nu p_{\Sigma}(g_{\mu \nu}-\chi_\mu \chi_\nu+u_\mu u_\nu) \delta(\chi)~=0~.
  \label{resulto}
\end{equation}
We recall that $n^\mu$ is the dust four-velocity, $u^\mu$ is the four-velocity of the shell on the boundary $\Sigma_1$, and $\chi^\mu$ is the spatial normal to the surface. The calculation can be performed using the results of Sec.\ref{secdynamicsthin}  for dust in the non-marginally bound case $\varepsilon = \text{const.} > 0$, but it also holds for null dust, as can be explicitly verified by applying Lemos' trick to \eqref{resulto}. 

The same applies to the diffeomorphism contribution:
\begin{align}
\mathcal{D}^\mathrm{shell}&=\frac{4 \pi r_*^2}{\sqrt{1+\varepsilon}} n^\mu h^\nu_r(g_{\mu \nu}-\chi_\mu \chi_\nu+u_\mu u_\nu) p_{\Sigma} \delta(\chi)= \notag \\
&=\frac{4 \pi r_*^2}{\sqrt{1+\varepsilon}} (n_r-n^\mu\chi_\mu \chi_r+n^\mu u_\mu u_r ) p_{\Sigma} \delta(\chi)=0   ~,
\end{align}
where $h_{\mu \nu} = g_{\mu \nu} + n_\mu n_\nu$ is the metric of the hypersurface orthogonal to the PG flow $n^\mu$, and in the third step we made use of the result \eqref{resulto}. This demonstrates that the tangential pressure does not contribute to the constraints, and consequently, that the presence of tangential pressure---along with a vanishing surface density $\sigma$---is compatible with the absence of a mass jump across the layer. 

To conclude, we evaluate the covariant conservation across the shell. The expression reads
\begin{equation} \label{DFs}
 D^a S_{ab}=-F_b^1+F_b^2   ~,
\end{equation}
where we must employ the induced metric $h_{ab}=\operatorname{diag}(-1, r_*^2, r_*^2 \sin^2 \theta)$ and the energy-momentum tensor
$S_{ab}=e^\mu_a e^\nu_b (h_{\mu \nu}+u_\mu u_\nu) p_{\Sigma}=p_{\Sigma}\cdot \operatorname{diag}(0, r_*^2, r_*^2 \sin^2 \theta)$. Instead
\begin{equation}
  F_a^{1,2}={^{(1,2)}}e^\mu_a  n^{(1,2)}_\nu n^{(1,2)}_\mu \chi^\nu_{(1,2)} \rho_{(1,2)} \,,
\end{equation}
is the matter flux, with:
\begin{align}
  &{^{(1,2)}}e^\mu_\tau=(\dot{T}_{(1,2)},\dot{r}_*,0,0)~,\\
  &{^{(1,2)}}e^\mu_\theta=(0,0,1,0) ~,\\
  &{^{(1,2)}}e^\mu_\phi=(0,0,0,1)~.
\end{align}
For both the spacetimes, we have $n_{\mu}^{(1,2)}=(-1, 0, 0, 0)$, and the only non-vanishing component of  $ D^a S_{ab}$ is 
\begin{equation}
     D^a S_{a \tau} = - 2 p_\Sigma  \frac{ \dot{r}_* }{r_*}~,
\end{equation}
while the only non-vanishing $F_{a}^{(1,2)}$ component is
\begin{equation}
    F^{(1,2)}_\tau = A_{(1,2)} \rho_{(1,2)} \dot{T}_{(1,2)}~,
\end{equation} 
with 
\begin{equation}
A_{(1,2)}=\frac{\dot{r}_*+N^r_{(1,2)}\dot{T}_{(1,2)}}{\sqrt{1+\varepsilon_{(1,2)}}}  \,, 
\end{equation}
being the time component of the normal $\chi^\mu$. We can then perform Lemos' trick, given by
\begin{align}
    & \varepsilon_{(1,2)}\rightarrow+\infty~, \\
    & \dot{T}_{(2)}= \sqrt{\varepsilon_2}\dot{V}_{(2)}-\frac{\dot{r}_*}{\sqrt{\varepsilon_2}}  ~, \\
  &  \Dot{T}_1=\sqrt{\varepsilon_1}\dot{U}_{(1)}-\frac{\dot{r}_*}{\sqrt{\varepsilon_1}} ~,
\end{align}
giving
\begin{equation}
 {2 p_{\Sigma} \dot{r}_*}=\frac{1}{4 \pi r_*}\left({\partial_{U_{(1)}} M(U_{(1)})}\dot{U}_{(1)}^2-{\partial_{V_{(2)}} M(V_{(2)})}\dot{V}_{(2)}^2 \right)\,.
 \label{b17}
\end{equation}
{We now specialize our black hole model to $f_\mathrm{qS}(r)$ and choose the corresponding $r_*$ surface to coincide with the one considered in Sec. \ref{sec_hawking_israel}. Recalling that $r_*=2GM(V_{(2)})=2GM(U_{(1)})$ {for $M>M_\mathrm{ext}$}, we obtain}
\begin{equation}
   \Dot{r}_*=2G \frac{\dd M}{\dd \tau}=2G \frac{\dd M}{\dd V_{(2)}}\Dot{V}_{(2)}=2G \frac{\dd M}{\dd U_{(1)}}\Dot{U}_{(1)} ~.
    \label{156} 
\end{equation}
Moreover, from the first Darmois-Israel junction condition we can easily write
\begin{align}
   &\dot{V}_{(2)}=\frac{1}{\frac{\gamma^2 \Delta}{(2GM(V_{(2)}))^2}-4G \frac{\dd M(V_{(2)})}{\dd V_{(2)}}} ~, \label{aaa} \\
  & \dot{U}_{(1)}=\frac{1}{\frac{\gamma^2 \Delta}{(2GM(U_{(1)}))^2}+4G \frac{\dd M(U_{(1)})}{\dd U_{(1)}}  } ~. \label{bbb}  
\end{align}
Putting all together, we get
\begin{widetext}
\begin{equation}
\begin{aligned}  p_\Sigma = \frac{1}{32 \left(\frac{\dd U_{(1)}}{\dd V_{(2)}}\right) \pi G^2 M} \left(4 \frac{\dd M}{\dd V_{(2)}} \frac{1}{\frac{\dd U_{(1)}}{\dd V_{(2)}}} G + \frac{\gamma^2 \Delta}{4 G^2 M^2}\right) \quad  \left[ \frac{\frac{1}{\frac{\dd U_{(1)}}{\dd V_{(2)}}}}{\left(4 \frac{\dd M}{\dd V_{(2)}} \frac{1}{\frac{\dd U_{(1)}}{\dd V_{(2)}}} G + \frac{\gamma^2 \Delta}{4 G^2 M^2}\right)^2} - \frac{1}{\left(4 \frac{\dd M}{\dd V_{(2)}} G - \frac{\gamma^2 \Delta}{4 G^2 M^2}\right)^2} \right] \,,
\end{aligned}
\end{equation}
\end{widetext}
{where $\dd U_{(1)}/\dd V_{2}$ is given in Eq. \eqref{jacob}.}
In the Schwinger case, after $M_\mathrm{ext}$ is reached, we have $\Dot{r}_*=0$, so eq. \eqref{b17} reduces {to $p_\Sigma=0$},
where we used \eqref{156} on the right-hand side.
\bibliography{sample}

\end{document}